\documentclass[aps,prx,superscriptaddress,nofootinbib,notitlepage,showpacs,floatfix]{revtex4-2}
\usepackage{algorithm}
\usepackage{algpseudocode}
\usepackage{hyperref}
\hypersetup{
    colorlinks=true, 
    linkcolor=cyan,
    citecolor=magenta, 
    filecolor=magenta, 
    urlcolor=cyan,
    runcolor=cyan
}

\usepackage{physics}
\usepackage{tikz}
\usetikzlibrary{positioning}
\def\checkmark{\tikz\fill[scale=0.4](0,.35) -- (.25,0) -- (1,.7) -- (.25,.15) -- cycle;}
\DeclareMathOperator{\sgn}{sgn}
\usepackage{physics}
\usepackage{accents}

\usepackage{xparse}
\ExplSyntaxOn
\NewDocumentCommand{\mref}{m}{\quinn_mref:n {#1}}
\seq_new:N \l_quinn_mref_seq
\cs_new:Npn \quinn_mref:n #1
 {
  \seq_set_split:Nnn \l_quinn_mref_seq { , } { #1 }
  \seq_pop_right:NN \l_quinn_mref_seq \l_tmpa_tl
  ( 
  \seq_map_inline:Nn \l_quinn_mref_seq
    { \ref{##1},\nobreakspace } 
  \exp_args:NV \ref \l_tmpa_tl 
  ) 
 }
\ExplSyntaxOff

\begin{document}
\title{HOQST: Hamiltonian Open Quantum System Toolkit}
\author{Huo Chen}
\email[]{huochen@usc.edu}
\affiliation{Department of Electrical and Computer Engineering, University of Southern California, Los Angeles, California 90089, USA}
\affiliation{Center for Quantum Information Science \& Technology, University of Southern California, Los Angeles, California 90089, USA}
\author{Daniel A. Lidar}
\affiliation{Department of Electrical and Computer Engineering, University of Southern California, Los Angeles, California 90089, USA}
\affiliation{Center for Quantum Information Science \& Technology, University of Southern California, Los Angeles, California 90089, USA}
\affiliation{Department of Chemistry, University of Southern California, Los Angeles, California 90089, USA}
\affiliation{Department of Physics and Astronomy, University of Southern California, Los Angeles, California 90089, USA}
\begin{abstract}
  \centerline{\textbf{Abstract}}
  We present an open-source software package called ``Hamiltonian Open Quantum System Toolkit" (HOQST), a collection of tools for the investigation of open quantum system dynamics in Hamiltonian quantum computing, including both quantum annealing and the gate-model of quantum computing. It features the key master equations (MEs) used in the field, suitable for describing the reduced system dynamics of an arbitrary time-dependent Hamiltonian with either weak or strong coupling to infinite-dimensional quantum baths. This includes the Redfield ME, the polaron-transformed Redfield ME, the adiabatic ME, the coarse-grained ME, and the universal Lindblad ME. HOQST also includes the stochastic Schr\"odinger equation with spin-fluctuators. We present an overview of the theories behind the various MEs and provide examples to illustrate typical workflows in HOQST. We present an example that shows that HOQST can provide order of magnitude speedups compared to QuTiP, for problems with time-dependent Hamiltonians. The package is ready to be deployed on high performance computing (HPC) clusters and is aimed at providing reliable open-system analysis tools for noisy intermediate-scale quantum (NISQ) devices. The HOQST Github repository \url{https://github.com/USCqserver/OpenQuantumTools.jl} provides the starting point for users. Detailed information can be found in the README file.
\end{abstract}

\maketitle
\section*{Introduction}
The theory of open quantum system has been an important subfield of quantum physics during the past decades with a rich collection of well-established methods~\cite{breuer_theory_2002,alicki_quantum_2007,weiss_quantum_2012}. Since perfect isolation of quantum systems is impossible, any quantum mechanical system must be treated as an open system in practice. The theory of open quantum system thus plays a major role in various applications of quantum physics, e.g., quantum optics~\cite{Gardiner:book,daley_quantum_2014}, quantum control~\cite{Wiseman:book}, and quantum computing (QC)~\cite{nielsen_quantum_2002}. It becomes even more relevant in the context of Hamiltonian quantum computing (HQC), broadly defined as analog QC performed via continuously and smoothly driven Hamiltonians, as opposed to discrete gate-model QC, where Hamiltonians are driven discontinuously. Well-known example of HQC include adiabatic quantum computing (AQC)~\cite{farhi_quantum_2000,albash_adiabatic_2018} and quantum annealing (QA)~\cite{kadowaki_quantum_1998, Hauke:2019aa}, as well as holonomic QC~\cite{HQC,Duan:2001ff}. For example, in QA the Hamiltonian needs to move continuously from the initial driver Hamiltonian to the final problem Hamiltonian, and therefore, unlike idealized gate-model quantum computers whose description often involves effective noise channels (completely positive maps), practical quantum annealers are better described by noise models derived directly from first principles~\cite{childs_robustness_2001,albash_decoherence_2015,amin_searching_2015,yip_quantum_2018, smirnov_theory_2018}. As the entire field of QC is now in the noisy intermediate-scale quantum (NISQ) era~\cite{preskill_quantum_2018}, an efficient and evolving framework of open quantum system simulation is essential for advancing our understanding of noise in quantum devices, as well for helping in the search for more effective error suppression and correction techniques~\cite{lidar_quantum_2013}. Moreover, the distinction between analog and discrete models of QC is to some degree arbitrary, since in reality, even gate-model QC involves continuous driving due to the finite bandwidth of signal generators and controllers. We thus view the gate-model of QC as part of HQC for the purposes of this work.

At present, there is an increasing number of software tools being developed for open system simulations. An important example of open-source software in this area is ``Quantum Toolbox in Python" (QuTiP)~\cite{johansson_qutip_2013}. It is one of the first packages in the field to adopt the modern software engineering paradigm and is actively maintained on Github since its release, with new features and enhancements being continuously added. However, since QuTiP is designed to be as general as possible, it lacks several tools and the computational performance required to address the new challenges we are now facing in the field of HQC.

Inspired by this challenge, and by the success of QuTiP, we present here a complementary and alternative open system simulation framework, which we call ``Hamiltonian Open Quantum System Toolkit" (HOQST). As the name suggests, the goal of HOQST is not to cover the entire field of open quantum system simulation but to focus on Hamiltonian QC, while retaining the flexibility to simulate systems subject to arbitrary time-dependent Hamiltonians. This focus gives us the ability to adopt domain-specific design choices and optimizations. The resulting implementation distinguishes itself from other available software by offering the following advantages:
\begin{itemize}
    \item HOQST is written in the Julia programming language~\cite{bezanson_julia:_2017}, which is designed for high performance computing.
    \item HOQST is built upon the ordinary differential equations (ODE) package \texttt{DifferentialEquations.jl}~\cite{rackauckas_differentialequations.jl_2017};
    thus HOQST also benefits from progress in the field of ODE solvers.
    \item Focusing solely on Hamiltonian QC, HOQST features several recently published master equations.
    \item HOQST includes tools that work beyond the weak coupling limit.
    \item HOQST provides a native interface for HPC clusters.
\end{itemize}

HOQST is developed following the Julia design philosophy: we intend it to be as user-friendly as possible without compromising performance. Although there is room for optimization, the first release of HOQST features reliable and efficient implementations of several key master equations (MEs) adopted in the HQC field, together with a highly modularized framework suitable for future development. Since the HOQST project started as an attempt to build a tool to simulate quantum annealing, it displays a certain bias towards QA. However, we reemphasize that it is broadly applicable to open quantum systems evolving subject to any time-dependent Hamiltonian. Besides the HOQST package itself, we provide error bounds on the MEs included. We also present examples to illustrate the typical workflows of HOQST. In particular, we focus on a three-qubit entanglement witness experiment performed using a quantum annealer~\cite{lanting_entanglement_2014}. Previous studies of open system models were unable to reproduce some of the key experimental features; we demonstrate that HOQST does now offer this capability.

\section*{Results}
\label{sec:results}

Throughout this work we consider a quantum mechanical system $\mathrm{S}$ coupled to a bath $\mathrm{B}$. The total Hamiltonian is assumed to have the following form
\begin{equation}\label{eq:full_hamiltonian}
    H = H_{\mathrm{S}} + H_{\mathrm{I}} + H_{\mathrm{B}} \ ,
\end{equation}
where $H_{\mathrm{S}}$ and $H_{\mathrm{B}}$ denote, respectively, the free system and bath Hamiltonians. $H_{\mathrm{I}}$ is the system-bath interaction, which is often written as
\begin{equation}\label{eq:interaction_hamil}
  H_\mathrm{I} = \sum_\alpha g_\alpha A_\alpha \otimes B_\alpha \ ,
\end{equation}
where $A_\alpha$ and $B_\alpha$ are dimensionless Hermitian operators acting on the system and the bath, respectively, and exclude both $I_{\mathrm{S}}$ and $I_{\mathrm{B}}$ (the identity operators on the system and the bath, respectively). The parameters $g_\alpha$ are sometimes absorbed into $B_\alpha$ but are kept explicit in HOQST, are have units of energy.
In addition, we assume for simplicity the factorized initial condition $\rho(0) = \rho_{\mathrm{S}}(0) \otimes \rho_{\mathrm{B}}$ for the joint system-bath state at the initial time $t=0$, where $\rho_\mathrm{B}$ is a Gibbs state at inverse temperature $\beta$
\begin{equation}
    \label{eq:rho_B}
    \rho_\mathrm{B} = \frac{e^{-\beta H_\mathrm{B}}}{\Tr[e^{-\beta H_\mathrm{B}}]} \ ,
\end{equation}
though we note that factorization is not necessary for a valid description of open system dynamics~\cite{Rodriguez:08,Dominy:2016xy}. We work in units of $\hbar=1$ and $k_B=1$ so that $\beta$ has units of inverse energy, or time.

Before proceeding, we transform the original Hamiltonian Eq.~\eqref{eq:full_hamiltonian} into a rotating frame defined by $U(t)$, i.e.
\begin{equation}
    \tilde{H}(t)=U^\dagger(t)HU(t) \ .
\end{equation}
The specific form of the unitary operator $U(t)$ will lead to different master equations and will be discussed in detail in subsequent sections. The goals of the rotation are to remove the pure-bath Hamiltonian $H_{\mathrm{B}}$ and to identify terms that remains small in different system bath coupling regimes. As long as $U(t)$ acts non-trivially on the bath, we may assume without loss of generality we  that the rotating frame Hamiltonian $\tilde{H}$ has the following form:
\begin{equation}
    \tilde{H} = \tilde{H}_{\mathrm{S}} + \tilde{H}_{\mathrm{I}} \ ,
\end{equation}
where $\tilde{H}_{\mathrm{S}}$ acts only the system and $\tilde{H}_{\mathrm{I}}$ acts jointly on the system and the bath. 
The Liouville von Neumann equation in this rotating frame is
\begin{equation}\label{eq:von_neumann}
    \frac{\partial}{\partial t} \tilde{\rho}(t) = -i[\tilde{H}(t),\rho(t)] \equiv \tilde{\mathcal{L}}(t)\rho(t) \ ,
\end{equation}
where $\tilde{\mathcal{L}}(t)$ denotes the Liouvillian superoperator. Once again we can always write
\begin{equation}\label{eq:interaction_hamil_tilde}
    \tilde{H}_\mathrm{I} = \sum_\alpha g_\alpha\tilde{A}_\alpha \otimes \tilde{B}_\alpha
\end{equation}  
where $\tilde{A}_\alpha$ and $\tilde{B}_\alpha$ are, respectively, system and bath operators (excluding identity). However, it is important to note that $\tilde{A}_\alpha$ and $\tilde{B}_\alpha$ do not necessarily correspond to  $U^\dagger(t)A_\alpha U(t)$ and $U^\dagger(t)B_\alpha U(t)$ in Eq.~\eqref{eq:interaction_hamil} because $U(t)$ may not preserve the tensor product structure (i.e., we allow for $U^\dagger A\otimes B U \neq U^\dagger A U \otimes U^\dagger B U$).

Let 
\begin{equation}
    \expval{\tilde{X}} = \Tr\left(\tilde{X} \tilde{\rho}_B \right)
\end{equation}
denote the expectation value of any rotating frame bath operator $\tilde{X}$ with respect to $\tilde{\rho}_B$. Then the two-point \emph{correlation function} is
\begin{equation}\label{eq:correlation}
    C_{\alpha\beta}(t_1, t_2) = g_\alpha g_\beta \expval{\tilde{B}_\alpha(t_1) \tilde{B}_\beta(t_2)} \ .
\end{equation}
If the correlation function is time-translation-invariant
\begin{equation}
    C_{\alpha\beta}(t_1, t_2) = C_{\alpha\beta}(t_1-t_2, 0) \equiv C_{\alpha\beta}(\tau)\ ,\quad \tau\equiv t_1-t_2\ ,
\end{equation}
then the \emph{noise spectrum} of the bath can be properly defined by taking the Fourier transform
\begin{equation}\label{eq:gamma}
  \gamma_{\alpha\beta}(\omega) = \int_{-\infty}^{\infty} C_{\alpha\beta}(\tau) e^{i\omega \tau}\dd{\omega} \ .
\end{equation}
The widely used Ohmic bath case is 
\begin{equation}
    \gamma^{\text{Ohmic}}_{\alpha\beta}(\omega) = 2\pi\eta g_\alpha g_\beta \frac{\omega e^{-\abs{\omega}/\omega_c}}{1-e^{-\beta\omega}} .
\label{eq:Ohmic-gamma}
\end{equation}

\subsection*{Timescales}
\label{sec:timescales}

We define the two timescales to measure the range of applicability~\cite{mozgunov_completely_2020}
\begin{equation}
  \label{eq:T1tauB-a}
  \frac{1}{\tau_{\mathrm{SB}}} = \int_0^{\infty}|C (\tau)|d\tau \ , \quad \tau_{\mathrm{B}} = \frac{\int_0^{t_f}\tau|C(\tau)|d\tau}{\int_0^{\infty}|C(\tau)|d\tau} \ .
\end{equation}
Here $t_f$ is the total evolution time, used as a cutoff which can often be taken as $\infty$. The quantity $\tau_{\mathrm{SB}}$ is the fastest system decoherence timescale, or timescale over which the system density matrix $\rho_S$ changes due to the coupling to the bath, in the interaction picture. The quantity $\tau_{\mathrm{B}}$ is the characteristic timescale of the decay of $C(\tau)$. Note that the expression for $\tau_{\mathrm{B}}$ becomes an identity if we choose $|C(\tau)| \propto e^{-\tau/\tau_{\mathrm{B}}}$ and take the limit $t_f\to\infty$. 

$\tau_{\mathrm{SB}}$ and $\tau_{\mathrm{B}}$ are the only two parameters relevant for determining the range of applicability of the various master equations discussed here~\cite{mozgunov_completely_2020}, with the Universal Lindblad Equation (ULE; see Methods) case being no exception~\cite{nathan_universal_2020}. For convenience we collect the corresponding error bounds here, before discussing the various MEs. Namely, the error bound of the Redfield master equation is
\begin{equation}
  \|\rho_{\text{true}}(t) -\rho_\mathrm{R}(t) \|_1 \le O\left(\frac{\tau_{\mathrm{B}}}{\tau_{\mathrm{SB}}}   e^{12t/\tau_{\mathrm{SB}}}\right)\text{ln}\left(\frac{\tau_{\mathrm{SB}}}{\tau_{\mathrm{B}}} \right)\ , 
  \label{err:Redfield}
\end{equation}
where $\rho_{\text{true}}(t)$ denotes the true (approximation-free) state.

The error bound of the Davies-Lindblad master equation is
\begin{equation}
  \|\rho_{\text{true}}(t) -\rho_\mathrm{D}(t) \|_1 \le  O\left( \left(\frac{\tau_{\mathrm{B}}}{\tau_{\mathrm{SB}}}  +\frac{1}{\sqrt{\tau_{\mathrm{SB}}\delta E}}\right)e^{12t/\tau_{\mathrm{SB}}}\right)\ ,
    \label{err:Lindblad}
\end{equation}
where $\delta E = $min$_{i\ne j}|E_i-E_j|$ is the level spacing, with $E_i$ the eigenenergies of the system Hamiltonian $H_\mathrm{S}$. This original version of this ME~\cite{davies_markovian_1974} does not directly allow for time-dependent driving, and we shall consider an adiabatic variant that does, the adiabatic master equation (AME)~\cite{albash_quantum_2012} (see Methods). The same error bound should apply in this case, since the difference is only in that the Lindblad operators are rotated in the AME case with the (adiabatically) changing eigenstates of $H_\mathrm{S}$, and the norms used to arrive at Eq.~\eqref{err:Lindblad} are invariant under this unitary transformation.

The error bound of the coarse-grained master equation (GCME; see Methods) is
\begin{align}
  \|\rho_{\text{true}}(t) -\rho_\mathrm{C}(t) \|_1 \le O\left(\sqrt{\frac{\tau_{\mathrm{B}}}{\tau_{\mathrm{SB}}}}e^{6t/\tau_{\mathrm{SB}}}\right) \ .
  \label{err:CGME}
\end{align}

The error of the polaron transform master equation (PTRE;  see Methods) can be separated into two parts. The first part comes from truncating the expansion to 2nd order. It can be bounded using the same expression as in Eq.~\eqref{err:Redfield}, with timescales defined by the polaron frame correlation $K(\tau)$:
\begin{align}
  \label{eq:Polaron_T1tauB}
      \frac{1}{\tau_{\mathrm{SB}}} = \Delta^2_m\int_0^{\infty}|K (\tau)|d\tau \ , \quad \tau_{\mathrm{B}} &= \frac{\int_0^{t_f}\tau|K(\tau)|d\tau}{\int_0^{\infty}|K(\tau)|d\tau} \ .
  \end{align}
A detailed discussion of the above quantities is presented in Supplementary Note 2 and 3 (the explicit form of $K(\tau)$ is also described in the Polaron Transform subsection in the Methods), and as far as we know the error bounds we derive here for the PTRE are new. We mention here that if the system-bath coupling strength $g_\alpha$ in Eq.~\eqref{eq:interaction_hamil} is sent to infinity, both $1/\tau_{\mathrm{SB}}$ and $\tau_{\mathrm{B}}/\tau_{\mathrm{SB}}$ go to $0$. Thus the PTRE works in the strong coupling regime.
The second part of the error is caused by ignoring the 1st and 2nd order inhomogeneous terms, which themselves are due to the polaron transformation breaking the factorized initial condition. We do not have a bound on this error yet, but numerical studies suggest it is small when $\rho_\mathrm{S}(0)$ is diagonal~\cite{jang_theory_2009}.

These bounds assume a Gaussian bath. For a non-Gaussian bath extra timescales relating to higher-order correlation functions generally appear, and the error bounds will contain additional terms.

\subsection*{Capabilities of HOQST and comparison with other quantum simulators}
\label{sec:capability}

Recent developments in the field of QC have led to an explosion of quantum software platforms~\cite{larose_overview_2019}, such as Qiskit~\cite{Qiskit, alexander_qiskit_2020}, pyQuil~\cite{smith_practical_2016} and ProjectQ~\cite{steiger_projectq_2018}.
Because some of these platforms include the capability to simulate noisy quantum circuits, we briefly compare their respective noise models and solver types in Table~\ref{table:noise_model}. Furthermore, for packages that support arbitrary time-dependent Hamiltonian and rely on MEs as solvers, we list their compatible MEs in Table~\ref{table:ME_support}.

\begin{table}[!htbp]
  \begin{center}
  \caption{Comparison chart for noise models and solver types adopted by different software packages. Here QC stands for quantum computing and ME stands for master equation. HOQST, QuTiP, Qiskit, Qiskit Pulse, pyQuil and ProjectQ are software package names.}
  \label{table:noise_model}
  \begin{tabular}{ l l l l l}
  \hline
    & HOQST  & QuTiP
    & Qiskit 
    & Qiskit Pulse
    \\
   \hline
   QC model & $t$-dependent Hamiltonian & $t$-dependent Hamiltonian & circuit & $t$-dependent Hamiltonian \\
   Noise model & system-bath coupling  & system-bath coupling & Kraus map& constant Lindblad operator \\
   Solver type & ME & ME & noisy gates & ME \\ 
   \hline\hline
   & pyQuil
   & ProjectQ
   \\
   \hline
   QC model  & circuit & circuit \\
    Noise model & Kraus map & Stochastic noise\\
   Solver type & Noisy gates & N/A
  \end{tabular}
  \end{center}
\end{table}

\begin{table}[!htbp]
  \begin{center}
  \caption{Comparison chart for master equation (ME) support. The table lists the supported ME types for HOQST, QuTiP and Qiskit Pulse. The abbreviation are: AME--adiabatic master equation; CGME--coarse-grained master equation; ULE--universal Lindblad equation; PTRE--polaron transformed Redfield equation. HOQST, QuTiP and Qiskit Pulse are software package names.}
  \label{table:ME_support}
  \begin{tabular}{ l l l l}
  \hline
    ME & HOQST  & QuTiP & Qiskit Pulse\\
   \hline
   constant Lindblad equation & \checkmark  & \checkmark & \checkmark \\
   Redfield equation & time/frequency form & frequency form & $\times$ \\
   AME ($t$-dependent Lindblad) & \checkmark & \checkmark & $\times$ \\
   CGME & \checkmark & $\times$ & $\times$ \\
   ULE & \checkmark & $\times$ & $\times$ \\
   PTRE & \checkmark & $\times$ & $\times$ \\
   Floquet-Markov formalisms & $\times$ & \checkmark & $\times$ \\
   Stochastic Schrodinger & spin-fluctuator & non-Hermitian effective Hamiltonian & $\times$
  \end{tabular}
  \end{center}
\end{table}

\begin{table}[!htbp]
  \begin{center}
      \caption{List of tutorials for HOQST.}
      \label{table:tutorials}
    \begin{tabular}{ l l}
      \hline
       Notebook & Description  \\
       \hline
       \hline
       Introductory \\
       \hline
       \texttt{01-closed\_system} & Introductory tutorial for solving closed system dynamics \\  
       \texttt{02-lindblad\_equation} & Time-independent Lindblad equation \\
       \texttt{03-single\_qubit\_ame} & Introductory open system simulation tutorial based on~\cite{albash_decoherence_2015} \\
       \texttt{04-polaron\_transformed\_redfield} & Polaron transformed Redfield equation [Eq.~\eqref{eq:PTRE}]\\
       \texttt{05-CGME\_ULE} & Coarse-grained ME and universal Lindblad equation [Eqs.~\eqref{eq:CGME} and~\eqref{eq:ULE}]\\
        \texttt{06-spin\_fluctuators} & Classical $1/f$ noise simulation [Eq.~\eqref{eq:classical_1_f}] \\
      \hline
      Advanced & \\
      \hline
      \texttt{hamiltonian/01-custom\_eigen} & Using user-defined eigendecomposition routine \\  
      \texttt{redfield/01-non\_positivity\_redfield} & An example of non-positivity in the Redfield equation \\
      \texttt{redfield/02-redfield\_multi\_axis\_noise} & Solving the Redfield equation with multi-axis noise\\
      \texttt{advanced/01-ame\_spin\_fluctuators} & Adiabatic master equation with spin-fluctuators [Eq.~\eqref{eq:hybrid_me}] \\
      \texttt{advanced/02-3\_qubit\_entanglement\_witness} & 3-qubit entanglement witness experiment
    \end{tabular}
  \end{center}
\end{table}

\begin{figure}[ht]
  \centering
\includegraphics[width=\textwidth]{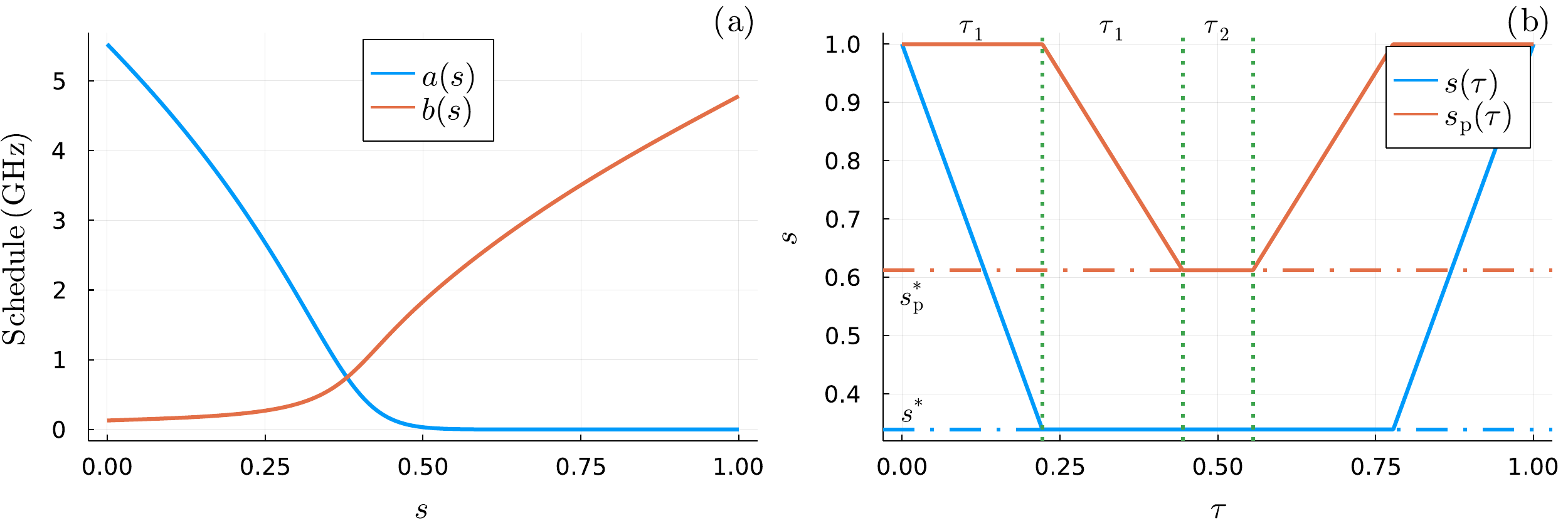}
  \caption{Annealing schedules and annealing parameters. In (a), we show the typical annealing schedules for the D-Wave device. In (b), we show the annealing parameters $s\pqty{\tau}$ and $s_\mathrm{p}\pqty{\tau}$ (see Eq.~\eqref{eq:3-qubit-witness}) used in the entanglement witness experiment. There are three stages in the experiment: i. first evolve $s(\tau)$ from $1$ to $s^{*}$ for $\tau\in[0,\tau_1]$ and then evolve $s_\mathrm{p}(\tau)$ from $1$ to $s_\mathrm{p}^*$ for $\tau\in[\tau_1,2\tau_1]$; ii. pause for a time $\tau_2$; iii. reverse the first stage. In our simulation, we choose $s^*=0.339$, $\tau_1 * t_f = 10\mu s$~\cite{albash_reexamination_2015} and $s_\mathrm{p}^*=0.612$ such that $2A(s_\mathrm{p}^*)\approx 1\mathrm{MHz}$~\cite{lanting_entanglement_2014}. The value of $\tau_2$ is varied to obtain the tunneling rate.}
  \label{fig:schedule}
\end{figure}

In this section, we benchmark the performance of HOQST against QuTiP. We consider only QuTiP because it is the single package in Table~\ref{table:noise_model} that provides a capability similar to HOQST, i.e., to simulate time-dependent open-system dynamics. The other packages all focus on the quantum-circuit model; thus, the comparison with HOQST is not meaningful. As a useful example, we choose the alternating-sectors-chain (ASC)~\cite{Mishra2018-ff} as the benchmark problem. The Hamiltonian of the $N$-qubit experiment is
\begin{equation}
  H_\mathrm{S}(\tau) = -a(\tau)\sum_{i=1}^N\sigma_i^x + b(\tau)H_{\mathrm{ASC}}
\end{equation}
where $\tau=t/t_f$ is the dimensionless time and $a(s)$ and $b(s)$ are the annealing schedules shown in Fig.~\ref{fig:schedule}(a). The alternating-sectors-chain Hamiltonian is
\begin{equation}
  H_\mathrm{ASC} = -\sum_{i=1}^{N-1}J_i\sigma_i^z\sigma_{i+1}^z \ ,
\end{equation}
where the coupling strength $J_i$ alternates between sectors  of size $n$
\begin{equation}
  J_i = \begin{cases}
    W_1 &\quad \textrm{if $\lceil i/n \rceil$ is odd} \\
    W_2 &\quad \textrm{otherwise}
  \end{cases} \ .
\end{equation}
To keep the problem size manageable, we fix $n=1$ and vary the system size $N$. The open-system model is given by
\begin{equation}
  H(\tau) = H_\mathrm{S}(\tau) +  g \sum_{i=1}^{N} \sigma^z_i\otimes B_i + H_\mathrm{B} \ ,
\end{equation}
where each qubit couples to an independent bath via $\sigma^z$ with equal coupling strength $g$, and $H_\mathrm{B}$ is the bath Hamiltonian. The bath is chosen to be Ohmic [Eq.~\eqref{eq:Ohmic-gamma}] with coupling strength $\eta g_\mathrm{S}^2 / \hbar ^2 = 1.2 \times 10^{-4}$, cutoff frequency $f_\mathrm{c} = 4\mathrm{GHz}$, and temperature $T = 12\mathrm{mK}$~\cite{Mishra2018-ff}. In the benchmark simulation, we solve the AME (frequency form Redfield equation) using both HOQST and QuTiP. Because of the large computational cost, the full AME simulation can hardly scale beyond a few qubits. HOQST provides interfaces to solve the AME in a low energy subspace. It can greatly speed up the computation if the evolution is confined within a small, low energy subspace. We also include this version of the AME solver in our benchmark (see Supplementary Note 6 for details). Finally, we ignore the Lamb shift in all the simulations since QuTiP does not include it (a significant drawback since in general the Lamb shift can have a strong effect~\cite{albash_quantum_2012}).

\begin{figure}[ht]
  \centering
  \includegraphics[width=0.6\textwidth]{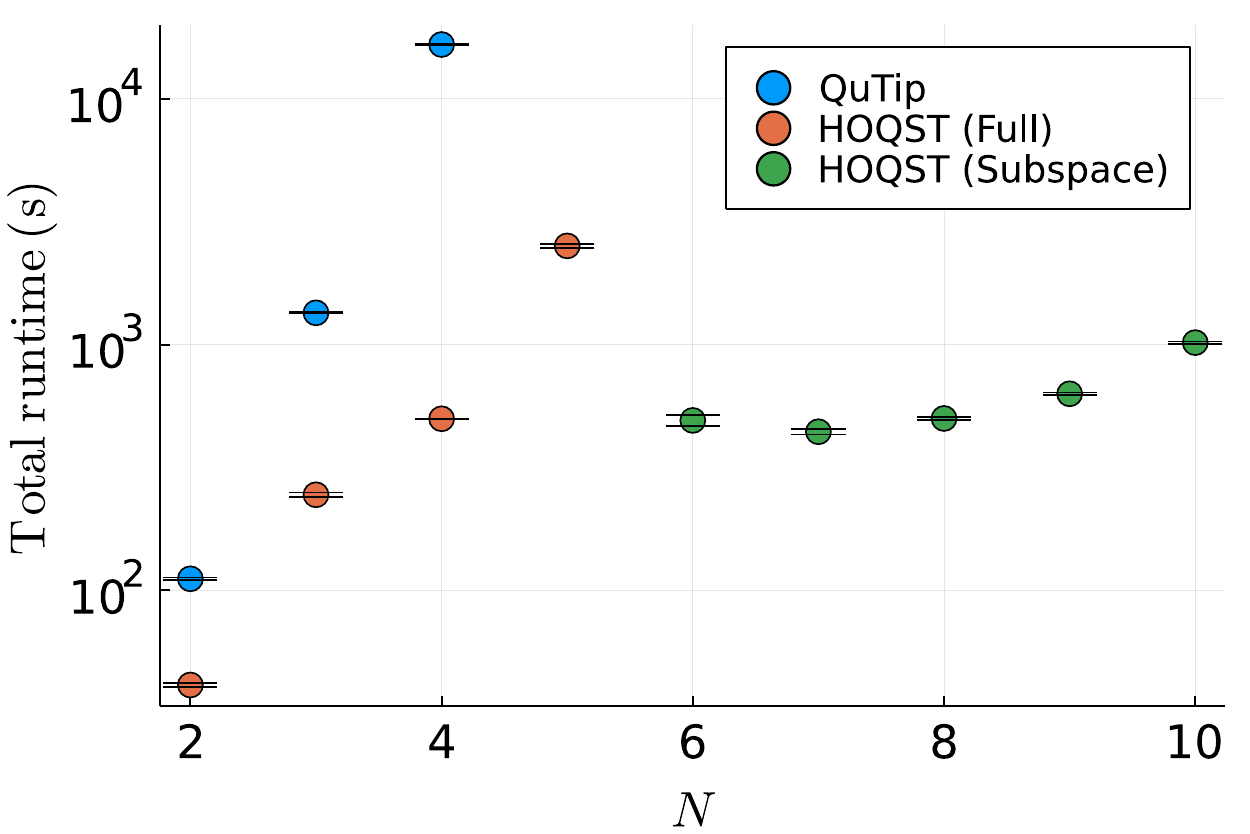}
  \caption{Total runtime \textit{vs} system size for the benchmark simulation of the alternating sectors chain problem. The x-axis ($N$) is the system size. Each data point corresponds to the average runtime of $7$-$10$ runs. Error bars represent 5 standard deviations. The ``QuTiP'' and ``HOQST (Full)'' points represent the runtime of full adiabatic master equation (AME) simulations using the corresponding packages. The ``HOQST (Subspace)'' points represent the runtime of AME simulation in the lowest 20-level subspace. The benchmark was done on a desktop computer with an Intel(R) Core(TM) i7-6700 @3.40GHz CPU and 16 GB memory. The software versions were QuTiP 4.6.2, HOQST 0.6.3, python 3.9.7 and Julia 1.6.3. The operating system was Ubuntu 20.04.3.}
  \label{fig:benchmark}
\end{figure}

The benchmark result is shown in Fig.~\ref{fig:benchmark}, where a significant runtime improvement of HOQST over QuTiP is observed. In fact, the QuTiP runtime became excessive for $N>4$, while HOQST remains nearly an order of magnitude faster even for $N=5$. Using HOQST's subspace truncation  capability allowed us to continue simulations up to $N=10$ without exceeding a runtime of $10^3$s. However, we note that we deliberately chose a benchmark problem that would demonstrate HOQST's advantage. There are areas of overlap between QuTiP and HOQST (e.g., the time-independent Lindblad equation) where the packages would perform similarly.

As a final remark, we emphasize that HOQST simulates the full open-system dynamics in the sense that the solution of the master equation is obtained up to a precision allowed by the underlying ODE algorithm. The computational cost of such a simulation scales exponentially with respect to the system size, and, without further assumptions or approximations, no algorithm with better scaling has been discovered. On the other hand, quantum Monte Carlo or tensor-network based algorithms could achieve superior scaling under additional assumptions or approximations. For example, if we know that the state stays within the space of matrix product state during the evolution, and the Liouvillian superoperator of the master equation could be effectively expressed in terms of matrix product operators, a tensor-network version of the ODE algorithm can solve the open-system dynamics efficiently. However, generically such assumptions are hard to satisfy, even approximately. Thus, it is not meaningful to benchmark HOQST against tensor-network based algorithm since the latter involves much stronger assumptions.

\subsection*{Entanglement witness experiment modeling}
HOQST has a large collection of tutorials located at a dedicated Github repo 
, which are summarized in Table~\ref{table:tutorials}.
As an illustrative yet non-trivial example, we next discuss the simulation of a three-qubit quantum annealing entanglement witness experiment. The entanglement witness experiment was proposed to provide evidence of entanglement in a D-Wave quantum annealing device~\cite{lanting_entanglement_2014}. An open system analysis of these experiments was performed using the AME~\cite{albash_reexamination_2015}, but failed to reproduce the observed width of the tunneling rate peaks. 
This is the impetus for us revisiting this experiment here. As we shall show, the new tools provided in HOQST allow us to much more closely match the experimental data than was possible before. 

\begin{figure}[ht]
  \center
  \includegraphics[width=0.3\textwidth]{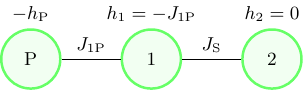}
  \caption{The Ising Hamiltonian of the 3-qubit entanglement witness experiment. The figure shows the graphical representation of $H_{\text{Ising}}$ in Eq.~\eqref{eq:3-qubit-witness}. Here each circle represents a qubit (qubit $1$, $2$ and an ancilla denoted as P). $h_i$ is the local field strength and $J_{ij}$ is the coupling strength between qubit $i$ and $j$. The Hamiltonian can be written as $\sum_i h_i\sigma^z_i + \sum_{ij}J_{ij}\sigma^z_i\sigma^z_j$. The goal of the experiment is to demonstrate entanglement between qubits 1 and 2, by probing the ancilla qubit. In our simulations, $h_1$, $h_2$, $J_{1\mathrm{P}}$ and $J_\mathrm{S}$ are fixed at $J_{1\mathrm{P}} = -h_1 = -1.8$, $J_\mathrm{S}=-2.5$, $h_2=0$.}
  \label{fig:3_qubit_coupling}
\end{figure}

The crux of the experiment is actually a form of tunneling spectroscopy~\cite{Berkley:2013bf}, where the goal is to find the energy gaps of the Hamiltonian $-a(s) \sum_{i} \sigma_{x}^{i}+b(s) H_{\mathrm{Ising}}$. This is done by observing the location of a peak in the tunneling rate as measured using a probe qubit.
The Hamiltonian of the 3-qubit-version of the experiment is
\begin{equation}\label{eq:3-qubit-witness}
  H_\mathrm{S}(\tau)=-a(s(\tau)) \sum_{i=1}^{2} \sigma^{x}_{i}-a(s_\mathrm{p}(\tau)) \sigma^{x}_{\mathrm{p}}+b(s(\tau)) H_{\mathrm{Ising}} \ ,
\end{equation}
where $a(s)$ and $b(s)$ are the annealing schedules, and $s(\tau)$ and $s_\mathrm{p}(\tau)$ are functions of the dimensionless time $\tau = t / t_f$, known as annealing parameters.
The Hamiltonian consists of two system qubits coupled to an ancilla system qubit, as shown in Fig.~\ref{fig:3_qubit_coupling}. The aforementioned location of the tunneling rate peak can be controlled by varying $h_\mathrm{P}$, and this information can be used to extract the energy gaps as a function of $s$. We refer interested readers to Refs.~\cite{lanting_entanglement_2014,albash_reexamination_2015} for more information.

The annealing schedules and annealing parameters are illustrated in Fig.~\ref{fig:schedule}. 
To extract the tunneling rate, we first perform the simulation with the initial all-one state $\ket{\psi(0)}=\ket{1}^{\otimes 3}$ for different $h_\mathrm{P}$ and $t_2 = \tau_2 t_f$ values. The population of the all-one state at the end of anneal is then obtained as a function of $h_\mathrm{P}$ and $\tau_2$: $P_{\ket{\mathbf{1}}}(h_\mathrm{P}, t_2) = \abs{\braket{\psi(t_f)}{\psi(0)}}^2 $. Lastly, we fit $P_{\ket{\mathbf{1}}}(h_\mathrm{P}, t_2)$ to the function $a e^{b t_2} + c e^{d t_2}$, from which the rate $\Gamma$ can be estimated
\begin{equation}
  \Gamma(h_\mathrm{P}) = -\pdv{P_{\ket{\mathbf{1}}}}{t_2}\bigg\vert_{t_2=0} = -ab - cd \ .
\end{equation}

For the open system model, we assume the qubits are coupled to independent baths~\cite{albash_reexamination_2015}
\begin{equation}
  \label{eq:open_sys_wit}
  H(t) = H_\mathrm{S}(t) +  \sum_{i=1}^{2}g_i\sigma^z_i\otimes B_i + g_\mathrm{P}\sigma_\mathrm{P}^z\otimes B_\mathrm{P} + H_\mathrm{B} \ ,
\end{equation}
but the bath coupling to the probe qubit $g_\mathrm{P}$ is much stronger than the coupling to the two system qubits $g_1=g_2=g_\mathrm{S}$.
In addition, we assume the bath is Ohmic [Eq.~\eqref{eq:Ohmic-gamma}] with coupling strength $\eta g_\mathrm{S}^2 / \hbar ^2 = 1.2732 \times 10^{-4}$, cutoff frequency $f_\mathrm{c} = 4\mathrm{GHz}$, and temperature $T = 12.5\mathrm{mK}$. We performed numerical simulations with different models of $B_\mathrm{P}$: 
\begin{enumerate}
    \item An Ohmic bath with interaction strength $g_\mathrm{P} = 10 g_\mathrm{S}$, using different flavors of the AME (see Methods, Sec. Adiabatic master equation).
    \item Hybrid Ohmic bath whose coupling strength to the Ohmic component is $g_\mathrm{P} = 10 g_\mathrm{S}$ and varying macroscopic resonant tunneling (MRT) width, using the PTRE (see the Polaron Transform subsection in the Methods).
\end{enumerate}

\begin{figure}[ht]
  \centering
  \includegraphics[width=0.6\textwidth]{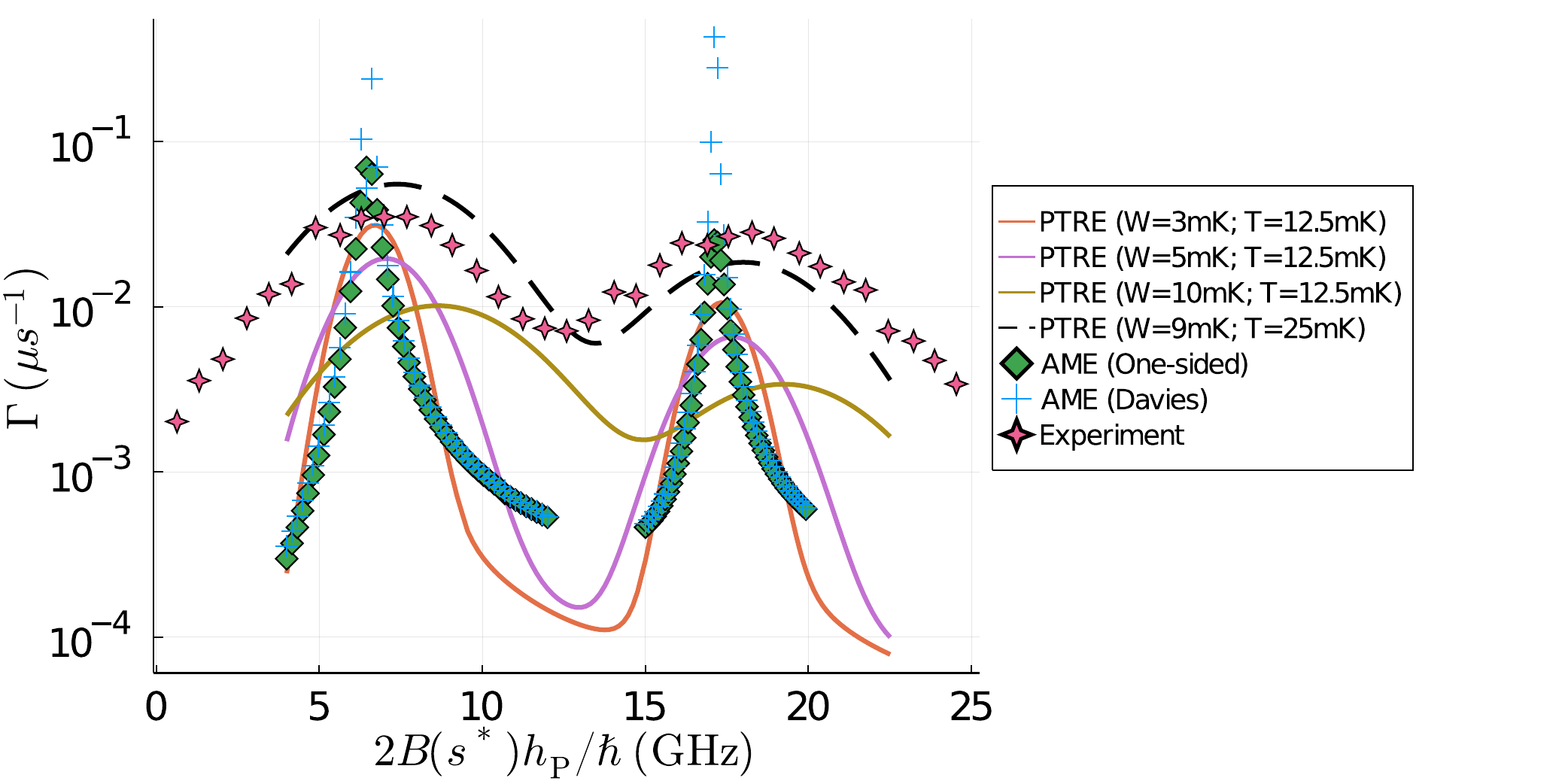}
  \caption{Tunneling rates obtained via different master equations compared with the experimental results. 
  The bath coupled to the system qubits is Ohmic with parameters: $\eta g_\mathrm{S}^2 / \hbar ^2 = 1.2732 \times 10^{-4}$, $f_\mathrm{c} = 4\mathrm{GHz}$ and $T = 12.5\mathrm{mK}$. The corresponding models and parameters of the bath coupled to the probe qubit used for different simulations are i. adiabatic master equation (AME): Ohmic with $g_\mathrm{P}=10g_\mathrm{S}$ and $s_\mathrm{p}^*=0.612$. ii. polaron transformed Redfield equation (PTRE): hybrid-Ohmic with $g_\mathrm{P}=10g_\mathrm{S}$ and various $W$, $T$ values. The cutoff frequency $f_\mathrm{c}$ is the same across different models. Here too $s_\mathrm{p}^*=0.612$. iii. PTRE with alternative parameters (black dashed line): hybrid-Ohmic with $W=9\mathrm{mK}$, $T=25\mathrm{mK}$, and $s_\mathrm{p}^*=0.59$, while the other parameters are the same as in case (2).}
  \label{fig:rate}
\end{figure}

The tunneling rates obtained via these different ME simulations are compared with the experimental results~\cite{lanting_entanglement_2014} in Fig.~\ref{fig:rate}. The reported experimental parameters are $T = 12.5\mathrm{mK}$ and $s_\mathrm{p}^*=0.612$~\cite{lanting_entanglement_2014}[see Fig.~\ref{fig:schedule}(b)]. 
Simulations using these parameters and two different flavors of the AME,  the one-sided AME [Eq.~\eqref{eq:one_sided_ame}] and the Lindblad form AME [Eq.~\eqref{eq:ame}], are plotted. The results demonstrate that these two AME flavors only differ significantly near the small gap region, but neither one matches the experimental results. No further improvement is observed by varying the AME parameters: the linewidth remains too narrow to match the experiment.

In contrast with the AME, the PTRE with $T = 12.5\mathrm{mK}$ and $s_\mathrm{p}^*=0.612$ exhibits a larger Gaussian linewidth broadening and closer agreement with the experimental data (solid curves in Fig.~\ref{fig:rate}). 
We note that if we increase $W$ while fixing $T$, the PTRE curve is stretched to the right. This is the result of the fluctuation-dissipation theorem, where $\varepsilon_\mathrm{L}$ scales quadratically with $W$. Such a shift can be compensated by increasing the temperature together with $W$ [see the black dashed curve in Fig.~\eqref{fig:rate}]. 

Despite the closer agreement using the PTRE at the reported experimental temperature and $s_\mathrm{p}^*$ values, there is still a mismatch between the theoretical and experimental amplitudes of the tunneling rate curves. A possible reason for this may be a discrepancy in the reported $A(s_\mathrm{p}^*)$ and its true value, attributable to annealing schedule fluctuations and integrated control errors in the early model of D-Wave annealer used in the experiment. To account for this, we performed PTRE simulation with different $T$ and $s_\mathrm{p}^*$ values from those reported~\cite{lanting_entanglement_2014}: $T=25\mathrm{mK}$ and $s_\mathrm{p}^*=0.59$ (we also set $W=9\mathrm{mK}$). The result is plotted as the black dashed line in Fig.~\ref{fig:rate} and shows significantly better agreement with the experimental results. This highlights the fact that the slow bath coupled to the probe qubit may have a different temperature than the Ohmic one. Moreover, this results illustrates the power of HOQST's range of ME implementations.

\section*{Conclusion}
\label{sec:discussion}

In conclusion, we presented a software package called Hamiltonian Open Quantum System Toolkit (HOQST). It is user-friendly and written in Julia. It supports various master equations with a wide joint range of applicability, as well as stochastic Hamiltonians to model $1/f$ noise. We demonstrated that HOQST can achieve order of magnitude speedups over QuTiP for problems with a time-dependent Hamiltonian. We also illustrated the use of HOQST in simulating open quantum system dynamics in a 3-qubit entanglement witness experiment. Whereas previous modeling of this experiment was unable to capture the reported linewidth, HOQST's implementation of the polaron-transformed Redfield equation (PTRE) was able to do so. We also derived new error bounds for the PTRE.

We expect HOQST to be useful for researchers working in the field of open quantum systems, dealing with systems governed by time-dependent Hamiltonians. HOQST provides both basic and advanced numerical simulation tools in this area, which can be applied to simulate superconducting qubits of all types, trapped ions, NV centers, silicon quantum dot qubits, etc. Future releases of HOQST will expand both the suite of open system models and range of quantum control and computation models it supports. 

\section*{Methods}
\label{sec:methods}
We present a brief review of the open system models and corresponding master equations supported by HOQST. We provide various additional technical details in the Supplementary Information.

\section*{Cumulant expansion}
The cumulant expansion is a technique originally designed for the perturbation expansion of stochastic differential equations~\cite{van_kampen_cumulant_1974}. This technique can be generalized to the open quantum system setting and allows a systematic description of the reduced system dynamics~\cite{breuer_theory_2002}. By applying this technique in different rotating frames, master equations with different ranges of applicability can be derived~\cite{breuer_theory_2002, smirnov_theory_2018,xu_non-canonical_2016}. Defining the projection operator $\mathcal{P}$
\begin{equation}
  \label{eq:projection_operator}
  \mathcal{P}\rho = \mathrm{tr}_\mathrm{B} \{\rho\} \otimes \rho_\mathrm{B}  \equiv \rho_\mathrm{S}\otimes \rho_\mathrm{B} \ ,
\end{equation}
the formal cumulant expansion of the Liouville von Neumann equation is
\begin{equation}\label{eq:formal_TCL}
  \frac{\partial}{\partial t} \mathcal{P}\tilde{\rho}(t) = \sum_n \mathcal{K}_n(t) \mathcal{P}\tilde{\rho}(t)\ ,
\end{equation}
where the nth order generator $\mathcal{K}_n(t)$ is
\begin{equation}\label{eq:cumulant_k}
  \mathcal{K}_n(t) = \int_0^t \dd{t_1} \int_0^{t_1}\dd{t_2} \cdots \int_0^{t_{n-2}}\dd{t_{n-1}}\expval{\tilde{\mathcal{L}}(t)\tilde{\mathcal{L}}(t_1)\cdots\tilde{\mathcal{L}}(t_{n-1})}_{\mathrm{oc}}\ ,
\end{equation}
and the quantities $\expval{\tilde{\mathcal{L}}(t)\tilde{\mathcal{L}}(t_1)\cdots\tilde{\mathcal{L}}(t_{n-1})}_{\mathrm{oc}}$ are known as ordered cumulants~\cite{breuer_theory_2002}. In HOQST, we consider only the first and second order generators, which are given by:
\begin{equation}
  \expval{\tilde{\mathcal{L}}(t)}_{oc} = \mathcal{P}\tilde{\mathcal{L}}(t)\mathcal{P} \ , \quad
  \expval{\tilde{\mathcal{L}}(t)\tilde{\mathcal{L}}(t_1)}_{oc} = \mathcal{P}\tilde{\mathcal{L}}(t)\tilde{\mathcal{L}}(t_1)\mathcal{P} \ .
\end{equation}
Incorporating higher order cumulants generally leads to more accurate results~\cite{breuer_theory_2002}.

\subsection*{Redfield equation}
The oldest and one of the most well-known MEs in this category is the Redfield equation~\cite{Redfield:66} (also known as TCL2~\cite{breuer_theory_2002}, where TCL$n$ stands for time-convolutionless at level $n$, arising from an expansion up to and including $\mathcal{K}_n(t)$), which directly follows from Eq.~\eqref{eq:formal_TCL} after choosing the rotation $U(t)$ to be
\begin{equation}
  \label{eq:US_UB}
      U(t) = U_\mathrm{S}(t)\otimes U_\mathrm{B}(t) \ , \quad U_\mathrm{S}(t)=T_+ \exp{-i \int_0^t H_\mathrm{S}(\tau)\dd{\tau}} \ , \quad  U_\mathrm{B}(t) = \exp{-iH_\mathrm{B}t} \ ,
  \end{equation}
where $T_+$ denotes the forward time-ordering operator.
After rotating back to the Schr\"{o}dinger picture, one of the most common forms of the Redfield equation is
\begin{equation}\label{eq:Redfield}
  \dot{\rho}_{\mathrm{S}}(t) = -i\comm{H_\mathrm{S}(t)}{\rho_\mathrm{S}(t)} - \sum_\alpha \comm{A_\alpha(t)}{\Lambda_\alpha(t)\rho_\mathrm{S}(t)} + h.c.
\end{equation}
where
\begin{equation}\label{eq:redfield_lambda}
  \Lambda_\alpha(t) = \sum_\beta \int_0^{t} C_{\alpha\beta}(t-\tau)U_\mathrm{S}(t, \tau)A_\beta(\tau)U_\mathrm{S}^\dagger(t, \tau) \dd{\tau} \ .
\end{equation}
This is the form used in HOQST. The error bound for the Redfield ME is given in Eq.~\eqref{err:Redfield}. We note that the current release of HOQST supports correlated baths for the Redfield and adiabatic master equation solvers. However, for simplicity, we henceforth focus on uncorrelated baths where $C_{\alpha\beta}(t)=\delta_{\alpha\beta}C_\alpha(t)$. 

The most significant drawback of the Redfield equation is the fact that it does not generate a completely-positive evolution, and in particular can result in unphysical negative states (density matrices with negative eigenvalues). Though formal fixes for this problem have been proposed~\cite{Gaspard:1999aa,Whitney:2008aa}, to address the issue in HOQST an optional positivity check routine is implemented at the code level, which can stop the solver if the density matrix become negative. In addition, three variants of Redfield equation that guarantee positivity, namely the adiabatic master equation (AME)~\cite{albash_quantum_2012}, the coarse-grained master equation (CGME)~\cite{mozgunov_completely_2020,majenz_coarse_2013}, and the universal Lindblad equation (ULE)~\cite{nathan_universal_2020}, are included in HOQST. We detail these MEs next.

\subsection*{Coarse-grained master equation (CGME)}
The CGME can be obtained from Eq.~\eqref{eq:Redfield} by first time-averaging the Redfield part, i.e., shifting $t \mapsto t+ t_1$ and applying $\frac{1}{T_a}\int_{-T_a/2}^{T_a/2}\dd{t_1}$. One then neglects a part of the integral to regain complete positivity. The result is~\cite{mozgunov_completely_2020}:
\begin{equation}\label{eq:CGME}
  \dot{\rho} = -i\comm{H_\mathrm{S}+H_{\mathrm{LS}}}{\rho} + \sum_\alpha \frac{1}{T_a}\int_{-T_a/2}^{T_a/2}\dd{t_1}\int_{-T_a/2}^{T_a/2}\dd{t_2} C_\alpha(t_2-t_1)\Big[A_\alpha(t+t_1)\rho_\mathrm{S}A^\dagger_\alpha(t+t_2) -\frac{1}{2}\acomm{A_\alpha(t+t_2) A_\alpha(t+t_1)}{\rho_\mathrm{S}}\Big] \ ,
\end{equation}
where $A_\alpha(t+t_1) = U^\dagger(t+t_1, t) A_\alpha(t) U(t+t_1, t)$ and the Lamb shift is given by
\begin{equation}
  H_{\mathrm{LS}} = \frac{i}{2T_a}\int_{-T_a/2}^{T_a/2}\dd{t_1}\int_{-T_a/2}^{T_a/2}\dd{t_2} \mathrm{sgn}(t_1-t_2) C_\alpha(t_2-t_1) A(t+t_2)A(t+t_1) \ .
\end{equation}
The quantity $T_a$ is the coarse-graining time, a phenomenological parameter that can be manually specified or automatically chosen based on the bath correlation function~\cite{majenz_coarse_2013}. HOQST uses a multidimensional h-adaptive algorithm~\cite{genz_remarks_1980} to perform the 2-dimensional integration. The error bound for the CGME is given in Eq.~\eqref{err:CGME}.

\subsection*{Universal Lindblad equation (ULE)}
The ULE~\cite{nathan_universal_2020} is a Lindblad-form master equation that shares the same error bound as the Redfield equation, i.e., Eq.~\eqref{err:Redfield}. A similar master equation with better accuracy, known as the geometric-arithmetic master equation (GAME),  can also be derived by using a different formula for the Lamb shift~\cite{Davidovic2020}. The formal form of the ULE is identical to the Lindblad equation:
\begin{equation}
  \label{eq:ULE}
      \dot{\rho}_{\mathrm{S}}(t) = -i\comm{H_\mathrm{S}(t)+H_{\mathrm{LS}}(t)}{\rho_\mathrm{S}(t)} + \sum_\alpha \Bigg[ L_\alpha(t)\rho L_\alpha^\dagger(t)-\frac{1}{2}\acomm{L_\alpha^\dagger(t)L_\alpha(t)}{\rho} \Bigg]\ ,
\end{equation}
where the time-dependent Lindblad operators are
\begin{equation}\label{eq:U-lindblad}
  L_\alpha(t) = \int_{-\infty}^{\infty} g_\alpha(t-\tau)U_\mathrm{S}(t,\tau)A_\alpha(\tau)U^\dagger(t,\tau) \dd{\tau} \ ,
\end{equation}
and the Lamb shift is
\begin{equation}\label{eq:U-lamb}
  H_\mathrm{LS}(t) = \sum_\alpha\frac{1}{2i}\int_{-\infty}^{\infty}\dd{s}\dd{s'}U(t, s)A_\alpha(s) g_\alpha(s-t)U(s,s')g_\alpha(t-s')A_\alpha(s')U^\dagger(t, s') \mathrm{sgn}(s-s') \ .
\end{equation}
In the above expression, $g_\alpha(t)$ is called the jump correlation and is the inverse Fourier transform of the square root of the noise spectrum
\begin{equation}
  g_\alpha(t) = \frac{1}{2\pi}\int_{-\infty}^{\infty}\sqrt{\gamma_\alpha(\omega)}e^{-i \omega t} \dd{\omega} \ .
\end{equation}
The integration limits of Eqs.~\eqref{eq:U-lindblad} and~\eqref{eq:U-lamb} are problematic in practice because the unitary $U_\mathrm{S}(t)$ does not go beyond $[0, t_f]$. In numerical implementation, we replace the integral limit with $\int_0^{t_f}$. This is a good approximation when $g(t)$ decays much faster than $t_f$. This is the form of ULE used in HOQST.

\subsection*{Adiabatic master equation (AME)}
\label{sec:ame}
To derive the AME, we replace $U_\mathrm{S}(t-\tau)$ in Eq.~\eqref{eq:redfield_lambda} with the ideal adiabatic evolution and apply the standard Markov assumption and the rotating wave approximation (RWA). The resulting equation is
\begin{equation}\label{eq:ame}
  \dot{\rho}_\mathrm{S}(t) = -i\comm{H_\mathrm{S}(t)+H_{\mathrm{LS}}(t)}{\rho_\mathrm{S}(t)} + \sum_{\alpha\beta}\sum_\omega\gamma_{\alpha\beta}(\omega)\bqty{L_{\omega,\beta}(t)\rho_\mathrm{S}(t)L^\dagger_{\omega,\alpha}(t)-\frac{1}{2}\acomm{L^\dagger_{\omega,\alpha}(t)L_{\omega,\beta}(t)}{\rho_\mathrm{S}(t)}} \ .
\end{equation}
The AME is in Davies form~\cite{davies_markovian_1974} and the Lindblad operators are defined by
\begin{equation}
  \label{eq:ame_lindblad}
  L_{\omega,\alpha}(t) = \sum_{\varepsilon_b -\varepsilon_a = \omega} \mel{\psi_a}{A_\alpha}{\psi_b}\dyad{\psi_a}{\psi_b}\ ,
\end{equation}
where $\varepsilon_{a}$ is the instantaneous energy of the $a$'th level of the system Hamiltonian, i.e., $H_\mathrm{S}(t)\ket{\psi_a(t)} =\varepsilon_{a}(t)\ket{\psi_a(t)}$. Finally, the Lamb shift term is
\begin{equation}
  H_{\mathrm{LS}}(t) = \sum_{\alpha\beta}\sum_{\omega} L^\dagger_{\omega,\alpha}(t) L_{\omega,\beta}(t)S_{\alpha\beta}(\omega) \ ,
\end{equation}
where 
\begin{equation}\label{eq:lambshift}
  S_{\alpha\beta}(\omega) = \frac{1}{2\pi}\int_{-\infty}^{+\infty} \gamma_{\alpha\beta}(\omega')\mathcal{P}(\frac{1}{\omega-\omega'})\dd{\omega'} ,
\end{equation}
with $\mathcal{P}$ denoting the Cauchy principal value. The error bound is given in Eq.~\eqref{err:Lindblad}. 

If the RWA is not applied, the resulting equation is called the one-sided AME: 
\begin{equation}\label{eq:one_sided_ame}
  \dot{\rho}_\mathrm{S}(t) = -i\comm{H_\mathrm{S}(t)}{\rho_\mathrm{S}(t)} + \sum_{\alpha\beta}\sum_\omega\Gamma_{\alpha\beta}(\omega)\comm{L_{\omega,\beta}(t)\rho_\mathrm{S}(t)}{A_\alpha} + h.c. \ ,
\end{equation}
where
\begin{equation}
  \label{eq:Gamma}
  \Gamma_{\alpha\beta}(\omega) = \int_0^{\infty} C_{\alpha\beta}(t)e^{i\omega t} \dd{t} = \frac{1}{2} \gamma_{\alpha \beta}(\omega)+\mathrm{i} S_{\alpha \beta}(\omega) \ .
\end{equation}
These two forms of the AME behave differently when the energy gaps are small because the RWA breaks down in such regions~\cite{mozgunov_completely_2020}. More importantly, like the Redfield equation, the one-sided AME does not generate a completely-positive evolution. The code-level positivity check routine works with this version of the AME as well.
\subsection*{Classical $1/f$ noise}
HOQST includes the ability to model $1/f$ noise, which is an important and dominant source of decoherence in most solid-state quantum NISQ platforms~~\cite{RevModPhys.86.361, Yip2021}, in particular those based on superconducting qubits~\cite{yan_flux_2016,nguyen_high-coherence_2019,quintana_observation_2017}. Fully quantum treatments of $1/f$ noise have been proposed~\cite{Bauernschmitt:1993aa,hsieh_unified_2018}, including for 	quantum annealing~\cite{smirnov_theory_2018}.
In HOQST we adopt the simpler approach of modeling $1/f$ noise as classical stochastic noise generated by a summation of telegraph processes, which has proved to be a good approximation to the fully quantum version~\cite{RevModPhys.86.361}. Specifically, we provide a quantum-trajectory simulation of the following stochastic Schr\"odinger equation
\begin{equation}\label{eq:classical_1_f}
  \ket{\dot{\Phi}} = -i \Big(H_\mathrm{S}+\sum_{\alpha} \delta_\alpha(t) A_\alpha\Big) \ket{\Phi} \ ,
\end{equation}
where each $\delta_\alpha(t)$ is a sum of telegraph processes
\begin{equation}\label{eq:summation_T}
  \delta_\alpha(t) = \sum_{i=1}^N T_i(t) \ ,
\end{equation}
where $T_i(t)$ switches randomly between $\pm b_i$ with rate $\gamma_i$. In the limit of $N\to\infty$ and $b_i\to\bar{b}$, if the $\gamma_i$'s are log-uniformly distributed in the interval $[\gamma_{\min},\gamma_{\max}]$ (with $\gamma_{\max} \gg \gamma_{\min}$), the noise spectrum of $\delta_\alpha(t)$ approaches a $1/f$ spectrum within the same interval~\cite{RevModPhys.86.361}. Empirically, we find that a good approximation can be achieved with relatively small $N$.

\subsection*{Hybrid model}
The most significant drawback of a purely classical noise model is that if its steady state is unique then it is the maximally mixed state. To see this, we first realize that each trajectory of Eq.~\eqref{eq:classical_1_f} generates a unitary acting on the space $\mathcal{S}(\mathcal{H}_S)$ of density matrices
\begin{equation}
  \mathcal{U}_k(t)\rho_\mathrm{S} = U_k(t) \rho_\mathrm{S} U_k^\dagger(t) \ .
\end{equation}
Averaging over the trajectories over a distribution $p(k)$ creates a unital (identity preserving) map from $\mathcal{S}(\mathcal{H}_S)$ into itself
\begin{equation}
  \label{eq:preserve_I}
      \bar{\mathcal{U}}(t) \rho_\mathrm{S}(0) = \int p(k)  \mathcal{U}_k(t)\rho_\mathrm{S} \dd{k}\ .
  \end{equation}
If the steady state $\rho_\infty$ is unique then we can define it as
\begin{equation}
  \label{eq:steady_state}
  \lim_{t\to \infty}\bar{\mathcal{U}}(t) \rho_\mathrm{S}(0) = \rho_\infty \quad \forall \rho_\mathrm{S}(0) \ .
\end{equation}
By unitality it would then follow that $\rho_\infty = {I}$, since we can choose $\rho_\mathrm{S}(0) = I$.

However, this is not what is observed in real devices, e.g., in experiments with superconducting flux~\cite{amin_searching_2015} or transmon~\cite{pokharel_demonstration_2018} qubits. To account for this, HOQST includes a hybrid classical-quantum noise model:
\begin{equation}\label{eq:hybrid_me}
  \dot{\rho} = -i \comm{H_\mathrm{S}+\sum_{\alpha} \delta_\alpha(t) A_\alpha}{\rho} +\mathcal{L}(\rho) \ ,
\end{equation}
where $\delta_\alpha(t)$ is the same random process as in Eq.~\eqref{eq:summation_T}, and $\mathcal{L}$ is the superoperator generated by the cumulant expansion~\eqref{eq:cumulant_k}. At present, HOQST supports the combination of $1/f$ noise with both the Redfield and adiabatic master equations.

\subsection*{Polaron transform}
\label{sec:polaron_transform}
If the bath operators in Eq.~\eqref{eq:interaction_hamil} are bosonic
\begin{equation}
  B_\alpha = \sum_{k} \lambda_{\alpha, k} (b^\dagger_{\alpha, k} + b_{\alpha, k}) \ , \qquad
  H_\mathrm{B} = \sum_{\alpha, k} \omega_{\alpha, k} b^\dagger_{\alpha,k}b_{\alpha, k} \ ,
\end{equation}
we can choose the joint system-bath unitary $U(t)$ in Eq.~\eqref{eq:US_UB} as~\cite{xu_non-canonical_2016}
\begin{equation}
  \label{eq:polaron_unitary}
  U_\mathrm{p}(t) = \exp{ - i \sum_{\alpha, k} A_\alpha^d \frac{g_{\alpha}\lambda_{\alpha,k}}{i\omega_{\alpha, k}}(b^\dagger_{\alpha, k} - b_{\alpha, k})} U_\mathrm{B}(t) \ ,
\end{equation}
where $A^{d}_\alpha$ is the diagonal component of $A_\alpha$ in the interaction Hamiltonian~\eqref{eq:interaction_hamil} and $U_\mathrm{B}(t)$ is given in Eq.~\eqref{eq:US_UB} (we use $\lambda$ instead of $g$ in $B_\alpha$ to distinguish it from the expansion parameter in Eq.~\eqref{eq:interaction_hamil_tilde}). The corresponding second order ME~\eqref{eq:Redfield} is known as the polaron-transformed Redfield equation (PTRE)~\cite{xu_non-canonical_2016} or the noninteracting-blip approximation (NIBA)~\cite{weiss_quantum_2012}. The PTRE has a different range of applicability than the previous MEs we have discussed. Whereas the latter apply under weak-coupling conditions, the transformation defined in Eq.~\eqref{eq:polaron_unitary} leads to a complementary range of applicability under strong-coupling. This particular form of Eq.~\eqref{eq:polaron_unitary} does not preserve the factored initial state, so that inhomogeneous terms are present after the transformation. However, if $\rho_\mathrm{S}(0)$ is diagonal then numerical studies of the effects of the inhomogeneous terms suggest that they can be ignored~\cite{jang_theory_2009}.

In addition, the PTRE can be extended beyond the spin-boson model by choosing a different form of the joint system-bath unitary~\eqref{eq:polaron_unitary}, as \cite{amin_macroscopic_2008, smirnov_theory_2018}
\begin{equation}\label{eq:general_polaron_unitary}
  U_\mathrm{p}(t) = U_\mathrm{B}(t) T_+\exp{-i\sum_\alpha A^{d}_\alpha g_{\alpha} \int_{0}^t B_\alpha(\tau)\dd{\tau}} \ .
\end{equation}
The two transformations in Eq.~\eqref{eq:polaron_unitary} and~\eqref{eq:general_polaron_unitary} lead to MEs with identical structure but slightly different expressions (see Supplementary Note 1 for details). Whether those differences make any physical significance is an interesting topic for further study. In this paper we choose to work with Eq.~\eqref{eq:polaron_unitary}.

Because the general form of the PTRE is unwieldy, we present its form for a standard quantum annealing model
\postdisplaypenalty=0
\begin{subequations}
  \begin{align}
      H(t) &= H_\mathrm{S}(t) + H_\mathrm{I} + H_\mathrm{B}\ , \quad H_\mathrm{I}= \sum_i g_i \sigma_i^z\otimes B_i  \\
      H_\mathrm{S}(t) & =a(t)H_\mathrm{driver} + b(t)H_\mathrm{prob}\ ,\label{eq:HS}
  \end{align}
\end{subequations}
where $a(t)$ and $b(t)$ are the annealing schedules, and $H_\mathrm{driver}$ and $H_\mathrm{prob}$ are the standard driver and problem Hamiltonians, respectively:
\begin{equation}
  H_\mathrm{driver} = -\sum_i \sigma^x_i\ , \quad H_\mathrm{prob} = \sum_i h_i \sigma_i^z + \sum_{i<j} J_{ij} \sigma_i^z \sigma_j^z \ ,
  \label{eq:Hann}
\end{equation}
where the Pauli matrix $\sigma^x$ acting on qubit $i$ is denoted by $\sigma^x_{i}$, etc.
The transformed Hamiltonian is
\begin{equation}
  \label{eq:polaron_hamil}
  \tilde{H}(t) =  a(t)\bigg[\sum_i \sigma^+_i \otimes \xi_i^+ (t) + \sigma^-_i\otimes \xi_i^- (t)\bigg]  +b(t)H_\mathrm{prob} \ , 
\end{equation}
where 
\begin{equation}
  \xi_i^{\pm}(t) =U^\dagger_{\mathrm{B}}(t) \exp{\pm \sum_k \frac{2g_i\lambda_{i,k}}{\omega_k}(b_{i,k}^\dagger - b_{i,k})} U_{\mathrm{B}}(t)\ .
\end{equation}
The Redfield equation corresponding to Eq.~\eqref{eq:polaron_hamil} is
\begin{equation}\label{eq:PTRE}
  \pdv{}{t}\tilde{\rho}_{\mathrm{S}}(t) = -i\comm{\tilde{H}_\mathrm{S}(t)+a(t)\sum_i\kappa_i\sigma_x}{\tilde{\rho}_\mathrm{S}(t)} - \sum_{i,\alpha} \comm{\sigma_i^\alpha}{\Lambda_i^\alpha(t)\tilde{\rho}_\mathrm{S}(t)} + h.c. \ ,
\end{equation}
where
\postdisplaypenalty=0
\begin{subequations}
  \begin{align}
      \label{eq:PTRE_lambda}
          \Lambda_i^\alpha(t) &= a(t)\sum_{\beta} \int_0^{t} a(\tau)K_i^{\alpha\beta}(t, \tau)\tilde{U}_\mathrm{S}(t, \tau)\sigma_i^\beta \tilde{U}_\mathrm{S}^\dagger(t, \tau) \dd{\tau} \\
          K_i^{\alpha\beta}(t, \tau) &=\expval{\xi_i^\alpha(t) \xi_i^\beta(\tau)}\\
          \kappa_i &= \expval{\xi_i^{\pm}(t)}
      \end{align}
\end{subequations}
and
\begin{equation}\label{eq:polaron_Hamiltonian}
  \tilde{U}_\mathrm{S}(t, \tau) = T_+\exp{-i\int_\tau^t \tilde{H}_\mathrm{S}(\tau') \dd{\tau'}} \ , \quad \tilde{H}_{\mathrm{S}}(t) = b(t)H_\mathrm{prob} \ .
\end{equation}
Here $K_i^{\alpha\beta}(t, \tau)$ is the two-point correlation function in the polaron frame [akin to the correlation function defined in Eq.~\eqref{eq:correlation}], and $\kappa_i$ corresponds to the first order cumulant generator in Eq.~\eqref{eq:cumulant_k} and is also known as the reorganization energy; it contributes a Lamb-shift-like term in Eq.~\eqref{eq:PTRE}. It is also worth mentioning that the polaron transformation~\eqref{eq:polaron_unitary} can be done partially, which means that in Eqs.~\eqref{eq:polaron_unitary} and~\eqref{eq:general_polaron_unitary}, $\alpha$ can be summed over a subset of system-bath coupling terms.

To solve this form of the PTRE in HOQST, the user can define a new correlation function $C_i^{\alpha\beta}(t,\tau) = a(t)a(\tau)K_{\alpha\beta}(t, \tau)$ and use the Redfield solver. An alternative approach is to make the Markov approximation in Eq.~\eqref{eq:PTRE_lambda}
\begin{equation}
  \int_0^t a(\tau) \cdots \dd{\tau} \to a(t)\int_0^\infty \cdots \dd{\tau}
\end{equation}
and write Eq.~\eqref{eq:PTRE} in Davies form~\cite{davies_markovian_1974} (see Supplementary Note 4 for more details). This leads to the same expression as the AME [Eq.~\eqref{eq:ame}], but with different Lindblad operators:
\begin{equation}\label{eq:ptre_lindblad}
  L_i^{\omega,\alpha}(t) = a(t)\sum_{\varepsilon_b -\varepsilon_a = \omega} \mel{\psi_a}{\sigma_i^\alpha}{\psi_b}\dyad{\psi_a}{\psi_b} \ ,
\end{equation}
where now $\ket{\psi_a}$ is the energy eigenstate of the Hamiltonian $\tilde{H}_\mathrm{S}(t)$, and the noise spectrum is
\begin{equation}
  \gamma_i^{\alpha\beta}(\omega) = \int_{-\infty}^{\infty}K_i^{\alpha\beta}(t)e^{i \omega t}\dd{t} \ .
\end{equation}
Then the AME solver can be used to solve this Lindblad-form PTRE.

For example, the following ME can be derived for the entanglement witness problem following the aforementioned procedure :
\begin{equation}
  \pdv{}{t}\tilde{\rho}_{\mathrm{S}}(t) = -i\comm{\tilde{H}_\mathrm{S}(t)+\tilde{H}_\mathrm{LS}(t)}{\tilde{\rho}_\mathrm{S}(t)} + \mathcal{L}_\mathrm{A}\Big[\tilde{\rho}_\mathrm{S}(t)\Big] + \mathcal{L}_\mathrm{P}\Big[\tilde{\rho}_\mathrm{S}(t)\Big] \ ,
\end{equation}
where
\begin{equation}
  \tilde{H}_\mathrm{S}(t) = -a(t) \sum_{i=1}^{2} \sigma^{x}_{i}+b(t) H_{\mathrm{Ising}} \ ,
\end{equation}
and the Liouville operators $\mathcal{L}_\mathrm{A}$ and $\mathcal{L}_\mathrm{P}$ corresponds to the AME part and PTRE part of this equation, respectively:
\postdisplaypenalty=0
\begin{subequations}
\begin{align}
  \mathcal{L}_\mathrm{A}(\rho) &= \sum_{i=1}^2\sum_\omega\gamma(\omega)\bqty{L_{\omega,i}(t)\rho L^\dagger_{\omega,i}(t)-\frac{1}{2}\acomm{L^\dagger_{\omega,i}(t)L_{\omega,i}(t)}{\rho}}\\
  \mathcal{L}_\mathrm{P}(\rho) &= \sum_{\alpha \in \Bqty{+,-}}\sum_\omega\gamma_\mathrm{P}(\omega)\bqty{L_\mathrm{P}^{\omega,\alpha}(t)\rho L_\mathrm{P}^{\omega,\alpha\dagger}(t)-\frac{1}{2}\acomm{L_\mathrm{P}^{\omega,\alpha\dagger}(t)L_\mathrm{P}^{\omega,\alpha}(t)}{\rho}} \ ,
\end{align}
\end{subequations}
where the Lindblad operators are defined in Eq.~\eqref{eq:ame_lindblad} and Eq.~\eqref{eq:ptre_lindblad} respectively. The function $\gamma(\omega)$ is the standard Ohmic spectrum and $\gamma_\mathrm{P}(\omega)$ is the polaron frame spectrum with a hybrid Ohmic form~\cite{amin_macroscopic_2008,smirnov_theory_2018} discussed in Supplementary Note 5. We provide the explicit form of $\gamma_\mathrm{P}(\omega)$ here
\begin{equation}
  \gamma_\mathrm{P}(\omega) = \int K(t) e^{i\omega t} \dd{t} = \int \frac{\dd{x}}{2\pi} G_\mathrm{L}(\omega-x)G_\mathrm{H}(x) \dd{x} \ ,
\end{equation}
where
\begin{equation}
  G_\mathrm{L}(\omega) = \sqrt{\frac{\pi}{2W^2}}\exp\Bigg[-\frac{(\omega-4\varepsilon_\mathrm{L})^2}{8W^2}\Bigg] \ ,
\end{equation}
and
\begin{equation}
  G_\mathrm{H}(\omega) = \frac{4\gamma(\omega)}{\omega^2 + 4\gamma(0)^2} \ .
\end{equation}
$G_\mathrm{L}(\omega)$ is the contribution of low frequency component, characterized by the MRT width $W$. Because $W$ and $\varepsilon_\mathrm{L}$ are connected through the fluctuation-dissipation theorem we have $W^2 = 2\varepsilon_\mathrm{L} T$; thus, hybridizing low frequency noise with an Ohmic bath introduces one additional parameter.

\subsection*{Numerical techniques}
\label{sec:numerical_technique}

\subsubsection*{Redfield backward integration}
To solve the Redfield or Redfield-like master equation~\eqref{eq:redfield_lambda}, one needs to integrate the unitary $U_{\mathrm{S}}$ backward in time at each ODE step. Such integrations are computationally expensive for long evolution times and become the bottleneck of the solver. To improve the efficiency of the solver, we introduce an additional parameter $T_a$ as the lower integration limit:
\begin{equation}
  \Lambda_\alpha(t) = \sum_\beta \int_{T_a}^{t} C_{\alpha\beta}(t-\tau)U_\mathrm{S}(t, \tau)A_\beta(\tau)U_\mathrm{S}^\dagger(t, \tau) \dd{\tau} \ .
\end{equation}
To justify this, note first that 
\begin{equation}
  \norm{\int_0^{T_a} C_{\alpha\beta}(t-\tau)U_\mathrm{S}(t, \tau)A_\beta(\tau)U_\mathrm{S}^\dagger(t, \tau) \dd{\tau}} \leq \int_{t-T_a}^{t} \abs{C(\tau')} \dd{\tau'} \ ,
  \label{eq:corr-bound}
\end{equation}
where $\norm{\cdot}$ is any unitarily invariant norm. To obtain this inequality, we perform a change of variable $t-\tau \to \tau'$ and make use of the fact that the operator $A_\beta(t)$ can always be normalized by absorbing a constant factor into the corresponding bath operator $B_\beta$. Second, note that in most applications the bath correlation function $C(\tau')$ decays fast compared with the total evolution time. As a result, the r.h.s. of Eq.~\eqref{eq:corr-bound} is small for sufficiently large $t$. The neglected part, i.e., the  integral over $[0,T_a]$, can thus be safely ignored so long as the r.h.s. of Eq.~\eqref{eq:corr-bound} is below the error tolerance of the numerical integration algorithm.

The same technique can also be applied to the ULE. The integration limits in Eqs.~\eqref{eq:U-lindblad} and~\eqref{eq:U-lamb} can be localized around $t$, i.e. replaced by $\int_{t-T_a}^{t+T_a}$. However, choosing an appropriate $T_a$ is a process of trial and error. The user needs to determine its value in a case-by-case manner.

\subsubsection*{Precomputing the Lamb shift}
Instead of evaluating the Lamb shift~\eqref{eq:lambshift} at each ODE step, to speed up the computations all the ME solvers support precomputing the Lamb shift on a predefined grid and use interpolation to fill up the values between the grid points.

\subsubsection*{Adiabatic frame}
For a typical annealing process, the total annealing time is usually much larger than the inverse energy scale of the problem
\begin{equation}
  t_f \gg \frac{1}{\min_{s\in{0,1}}\big[\max(A(s), B(s))\big]} \ .
\end{equation}
Informally, the frequency of the oscillation between the real and imaginary part of the off-diagonal elements of $\rho_\mathrm{S}$ in the neighborhood of $s$ is positively proportional to both $A(s)$ and $B(s)$. As a consequence, directly solving the dynamics in the Schr\"{o}dinger picture is challenging because the algorithm needs to deal with the fast oscillations induced by the Hamiltonian, thus impacting the step size. HOQST includes an optional pre-processing step to rotate the Hamiltonian into the adiabatic frame~\cite{klarsfeld_magnus_1992}. If the evolution is in the adiabatic limit, the off-diagonal elements of the density matrix in this frame should approximately vanish. The fast oscillation is absent and a large step size can be taken by the ODE solver. This technique provides advantages if the user wishes to repeatedly solve the same problem with different parameters. See Supplement Method 1: Adiabatic frame for a brief summary.

\subsubsection*{Quantum trajectories method}
HOQST implements a quantum-trajectory solver for the AME~\cite{yip_quantum_2018}. Using the native distributed memory parallel computing interface of both \texttt{Julia} and \texttt{DifferentialEquations.jl}, the quantum-trajectory simulations can take advantage of HPC clusters with minimum changes in the code. In addition, classical $1/f$ noise can be infused into the AME trajectory solver to generate the hybrid dynamics described in Eq.~\eqref{eq:hybrid_me}.

\section*{Acknowledgments}
The authors are grateful to Jenia Mozgunov, Tameem Albash, Humberto Munoz Bauza, Ka Wa Yip and Vinay Tripathi for useful discussions and feedback. The authors also thanks Grace Chen for the HOQST logo design. This research is based upon work supported by the Office of
the Director of National Intelligence (ODNI), Intelligence Advanced
Research Projects Activity (IARPA) and the Defense Advanced Research Projects Agency (DARPA), via the U.S. Army Research Office contract W911NF-17-C-0050, and by the National Science Foundation the Quantum Leap Big Idea under Grant No. OMA-1936388. The views and conclusions contained herein are those of the authors and should not be interpreted as necessarily
representing the official policies or endorsements, either expressed or
implied, of the ODNI, IARPA, DARPA, ARO, or the U.S. Government. The U.S. Government
is authorized to reproduce and distribute reprints for Governmental
purposes notwithstanding any copyright annotation thereon.
The authors acknowledge the Center for Advanced Research Computing (CARC) at the University of Southern California for providing computing resources that have contributed to the research results reported within this publication. URL: \url{https://carc.usc.edu}.
\appendix
\section*{SUPPLEMENTARY NOTE 1: THE POLARON TRANSFORMATION}
\label{sec:formalism_polaron}

\subsection*
{Two different formulations}

We illustrate two different formulations of the polaron transformation via the simplest Hamiltonian 
\begin{align}
\label{eq:polaron_example_H}
    H &= H_\mathrm{S}\otimes I_B + H_{\mathrm{SB}} + I_S\otimes H_\mathrm{B} \notag\\
    &\equiv (\varepsilon\sigma_z + \Delta\sigma_x)\otimes I_B + g\sigma_z\otimes \sum_k \lambda_k(b_k^\dagger + b_k) + I_S\otimes\left(\sum_k \omega_k b_k^\dagger b_k\right)  \ .
\end{align}
The standard polaron transformation rotates this Hamiltonian via the unitary~\cite{xu_non-canonical_2016}
\begin{equation}
    \label{eq:app_polaron_unitary}
    U_p = \exp{- i\sigma_z \sum_k \frac{g\lambda_k}{i\omega_k}(b^\dagger_k - b_k)}
\end{equation}
which leads to the interaction picture Hamiltonian
\begin{equation}
    \tilde{H} = \varepsilon\sigma_z + \Delta\xi_+\sigma_+ + \Delta\xi_-\sigma_- + H_\mathrm{B} \ ,
\end{equation}
where
\begin{equation}
    \xi_\pm = \exp{\pm 2i \sum_k \frac{g\lambda_k}{i\omega_k}(b^\dagger_k - b_k)} \ .
\end{equation}
The only quantities that matter in the 2nd order TCL formalism are the average and two-point correlation function of $\xi_\pm(t)$ given by

\begin{subequations}
    \begin{align}
        \label{eq:polaron_k1}
        \expval{\xi_{\pm}(t)} &= \Tr_\mathrm{B}\{U^\dagger_\mathrm{B}(t) \xi_\pm U_\mathrm{B}(t)\rho_\mathrm{B}\} = \Tr_\mathrm{B}\{ \xi_\pm \rho_\mathrm{B}\}\\
        \label{eq:polaron_k2}
        \expval{\xi_\alpha(t)\xi_\beta(0)} &=\Tr_\mathrm{B}\{U^\dagger_\mathrm{B}(t) \xi_\alpha U_\mathrm{B}(t) \xi_\beta \rho_\mathrm{B}\} \ ,
    \end{align}
\end{subequations}
where $\alpha, \beta \in \{+, -\}$ and $U_\mathrm{B}(t) = \exp{-iH_\mathrm{B} t}$. To arrive at the second equality in Eq.~\eqref{eq:polaron_k1} we used $[\rho_\mathrm{B},U_\mathrm{B}]=0$, which is true for $\rho_\mathrm{B}$ in a Gibbs state.

An alternative approach~\cite{smirnov_theory_2018} is to first rotate the Hamiltonian~\eqref{eq:polaron_example_H} w.r.t. $H_\mathrm{B}$ and then rotate it again by
\begin{equation}
        \label{eq:app_general_polaron_unitary}
        U_d(t) = T_+\exp{-i \sigma_z \int_{0}^t B(\tau)\dd{\tau}} \ ,
\end{equation}
where 
\begin{equation}
    B(\tau) = U^\dagger_\mathrm{B}(\tau) g\sum_k \lambda_k(b_k^\dagger + b_k) U_\mathrm{B}(\tau) \ .
\end{equation}
It is worth noting that in this formalism the bath does not need to be bosonic. Let us define 
\begin{equation}
    \Xi_\pm(t) = T_+\exp{\pm i \int_{0}^t B(\tau) \dd{\tau}} \ .
\end{equation} 
Then the effective Hamiltonian in this two-fold interaction picture is 
\begin{equation}
    \tilde{H} = \varepsilon\sigma_z + \Delta\zeta_+(t)\sigma_+ + \Delta\zeta_-(t)\sigma_- \ ,
\end{equation}
where
\begin{equation}
    \zeta_\pm(t) = \Xi^\dagger_\mp(t) \Xi_\pm(t) \ .
\end{equation}
Again, the 2nd order ME only depends on
\begin{subequations}
    \begin{align}
        \label{eq:general_polaron_k1}
        \expval{\zeta_{\pm}(t)} &= \Tr_\mathrm{B}\{\Xi^\dagger_\mp(t) \Xi_\pm(t)\rho_\mathrm{B}\} \\
        \label{eq:general_polaron_k2}
        \expval{\zeta_\alpha(t_1)\zeta_\beta(t_2)} &=\Tr_\mathrm{B}\{\Xi^\dagger_{\bar{\alpha}}(t_1) \Xi_\alpha(t_1)\Xi^\dagger_{\bar{\beta}}(t_2) \Xi_\beta(t_2)\rho_\mathrm{B}\} \ ,
    \end{align}
\end{subequations}
where $\bar{\alpha}$ means the opposite symbol of $\alpha$ in $\{+,-\}$. Appendices D.4 and D.5 in \cite{smirnov_theory_2018} give detailed calculations for Eq.~\eqref{eq:general_polaron_k1} and~\eqref{eq:general_polaron_k2}.

\subsection*{Correlation function in polaron frame}
\label{sec:formulas}

We now discuss Eqs.~\mref{eq:polaron_k1, eq:polaron_k2,eq:general_polaron_k1, eq:general_polaron_k2} for the spin-boson model. We first introduce the bath spectral function~\cite{leggett_dynamics_1987}
\begin{equation}
    J(\omega) = \sum_k \lambda^2_k \delta(\omega-\omega_k) \ ,
\end{equation}
which is usually taken into the continuous limit. The standard result~\cite{leggett_dynamics_1987,lee_accuracy_2012} for Eqs.~\eqref{eq:polaron_k1} and~\eqref{eq:polaron_k2} are
\begin{equation}
    \label{eq:kappa}
    \kappa = \expval{\xi_{\pm}(t)} = \exp{-2g^2\sum_k\frac{\lambda_k^2}{\omega_k^2}\coth(\beta\omega_k/2)} = \exp{-2g^2\int_0^\infty\frac{J(\omega)}{\omega^2}\coth(\beta\omega/2)\dd{\omega}} \ ,
\end{equation}
and
\begin{subequations}
    \begin{align}
        \expval{\xi_\alpha(t)\xi_\alpha(0)} &= \kappa^2e^{-\phi(t)} \\
        \expval{\xi_\alpha(t)\xi_\beta(0)}&=\kappa^2e^{\phi(t)} \ ,
    \end{align}
\end{subequations}
where
\begin{equation}\label{app:Q_phi}
    \phi(t)=4g^2\int_0^\infty\dd{\omega}\frac{J(\omega)}{\omega^2}\Big(\coth(\frac{\beta\omega}{2})\cos(\omega t)-i\sin(\omega t)\Big) \ .
\end{equation}
For spectral functions which lead to $\kappa \to 0$ and $\phi(t) \to \infty$ (e.g., an Ohmic bath), $\kappa^2 e^{\phi(t)}$ converges to a finite number. Thus the standard result for an Ohmic bath is
\begin{subequations}
    \begin{align}
        \label{eq:polaron_avg_ohmic}
        \expval{\xi_+(t)\xi_+(0)} &= \expval{\xi_-(t)\xi_-(0)} = 0\\
        \label{eq:polaron_correlation_ohmic}
        \expval{\xi_+(t)\xi_-(0)} &= \expval{\xi_-(t)\xi_+(0)} = \exp(-4Q_2(t)-4iQ_1(t)) \ ,
    \end{align}
\end{subequations}
where $Q_2(t)$ and $Q_1(t)$~\cite{leggett_dynamics_1987} are
\begin{subequations}
    \begin{align}
        Q_1(t) &\equiv g^2\int_0^\infty \frac{J(\omega)}{\omega^2} \sin(\omega t) \dd{\omega} \\
        Q_2(t) &\equiv g^2\int_0^\infty \frac{J(\omega)(1-\cos(\omega t))}{\omega^2}\coth(\beta\hbar\omega/2) \dd{\omega} \ .
    \end{align}
\end{subequations}
By substituting the Ohmic spectral function $J(\omega) = \eta \omega e^{\omega/\omega_c}$, the explicit form of $Q_1$ and $Q_2$ can be calculated. Here we take another approach by noticing that
\begin{equation}
    -Q_2(t)-iQ_1(t) = \int_0^t\int_{-\infty}^{0} C(t_1, t_2) \dd{t_1} \dd{t_2}\ ,
\end{equation}
where $C(t_1, t_2)$ is the bath correlation function (defined in Eq.~(9) of the main text) that can also be expressed in terms of the spectral function
\begin{equation}
    C(t_1, t_2) = g^2\int_0^{\infty}\dd{\omega} J(\omega)\coth(\beta\omega/2)\cos(\omega (t_1-t_2)) - iJ(\omega)\sin(\omega(t_1-t_2)) \ .
\end{equation}
Thus Eq.~\eqref{eq:polaron_correlation_ohmic} can be rewritten as
\begin{equation}\label{eq:app_polaron_correlation}
    \expval{\xi_+(t)\xi_-(0)} = \expval{\xi_-(t)\xi_+(0)} = \exp{4\int_{-\infty}^{\infty} \gamma(\omega) \frac{e^{-i\omega t}-1}{\omega^2}\dd{\omega}} \ ,
\end{equation}
where $\gamma(\omega)$ is the spectral density defined in Eq.~(11) of the the main text.

For Eq.~\eqref{eq:general_polaron_k1} and~\eqref{eq:general_polaron_k2}, we follow Ref.~\cite{smirnov_theory_2018} and write down the the result here for comparison. The 1st order average is
\begin{equation}
    \expval{\zeta_{\pm}(t)} = \exp{-4Q_2(t)-4iQ_1(t)} \ ,
\end{equation}
which is supposed to decrease exponentially fast for $t>0$. The two-point correlation functions are
\begin{subequations}
    \begin{align}
        \label{eq:general_polaron_ohmic}
        \expval{\zeta_+(t)\zeta_+(0)} &= \expval{\zeta_-(t)\zeta_-(0)} = 0\\
        \label{eq:general_polaron_correlation_ohmic}
        \expval{\zeta_+(t_1)\zeta_-(t_2)} &= \expval{\zeta_-(t_1)\zeta_+(t_2)} = \exp\bigg\{-4Q_2(t)-4iQ_1(t) +4i\big[C(t_1)-C(t_2)\big]\bigg\} \ .
    \end{align}
\end{subequations}
The difference between Eqs.~\eqref{eq:polaron_correlation_ohmic} and~\eqref{eq:general_polaron_correlation_ohmic} are easily noticeable. The physical origins and significance of this difference is a subject for future studies.

\section*{SUPPLEMENTARY NOTE 2: ERROR ANALYSIS}
\label{sec:PTRE_error}

We present an error analysis based on the standard form of the polaron transformation~\eqref{eq:app_polaron_unitary}. Our strategy is to divide the error into separate parts and use Lemma 1 in Ref.~\cite{mozgunov_completely_2020} to estimate the total error of the master equation. Besides those already presented in~\cite{mozgunov_completely_2020}, there are two additional error terms in the PTRE, the physical origin of which are the inhomogeneous terms resulting from the breakdown of the factorized initial condition by $U_p$:
\begin{equation}
    U_p^\dagger \rho_\mathrm{S}(0)\otimes \rho_\mathrm{B} U_p \neq \tilde{\rho}_\mathrm{S}(0) \otimes \tilde{\rho}_\mathrm{B} \ .
\end{equation}
In addition to the projector defined in Eq. (21) in the main text, we define its complementary $\mathcal{Q} \equiv I -\mathcal{P}$. The standard projection operator technique~\cite{breuer_theory_2002} leads to
\begin{equation}\label{eq:formal_TCL_inhomogeneous}
    \frac{\partial}{\partial t} \mathcal{P}\tilde{\rho}(t) = \sum_n \mathcal{K}_n(t) \mathcal{P}\tilde{\rho}(t) + \sum_n \mathcal{I}_n(t)\mathcal{Q}\tilde{\rho}(0)\ ,
\end{equation}
where the rightmost expression is the contribution of the inhomogeneous terms. Ref.~\cite{chang_non-markovian_1993} calculates $\mathcal{I}_n(t)$ up to $n=4$ and shows they are identical to those for $\mathcal{K}_n(t)$ except that the final $\mathcal{P}$ in each term is replaced by a $\mathcal{Q}$. Here we conjecture that this is also true for $n>4$, and then use the same error bound as in Ref.~\cite{mozgunov_completely_2020} (called the Born approximation error there), to bound the error of truncating to the order of $\mathcal{I}_2$:
\begin{equation}
    \label{eq:inhomogeneous_error_2}
    \mathcal{E}^{(\infty)}_I = \norm{\sum_{n=1}^{\infty} \mathcal{I}_n(t)\mathcal{Q}\tilde{\rho}(0) - \sum_{n=1}^{2} \mathcal{I}_n(t)\mathcal{Q}\tilde{\rho}(0)}_1 \ .
\end{equation}
To simplify the notation, we define $\rho_{\text{true}}(t)$ as the solution of Eq.~\eqref{eq:formal_TCL_inhomogeneous} when $n \to \infty$. The PTRE error bound can be written as the sum of three terms:
\begin{equation}
    \|\rho_{\text{true}}(t) -\rho_{\text{PTRE}}(t) \|_1 \leq \mathcal{E}_{BM} + \mathcal{E}_{I}^{(\infty)} + \mathcal{E}_{I}^{(2)} \ ,
\end{equation}
where $\mathcal{E}_{BM}$ is the error bound presented in Ref.~\cite{mozgunov_completely_2020}:\footnote{The error bound in Ref.~\cite{mozgunov_completely_2020} includes the error due to changing $\int^{t} \rightarrow \int^{\infty}$ in the Redfield equation. We employ Eq.~\eqref{eq:bound_BM} despite not making this approximation, because the upper bound remains valid without it.}
\begin{equation}
    \label{eq:bound_BM}
    \mathcal{E}_{BM} \leq O\left(e^{\frac{12 t}{\tau_{S B}}} \frac{\tau_{B}}{\tau_{S B}}\right) \ln \frac{\tau_{S B}}{\tau_{B}} \ ,
\end{equation}
$\mathcal{E}_I^{(\infty)}$ is given in Eq.~\eqref{eq:inhomogeneous_error_2} and $\mathcal{E}^{(2)}_I=\norm{\sum_{n=1}^{2} \mathcal{I}_n(t)\mathcal{Q}\tilde{\rho}(0)}_1$. Here we argue that $\mathcal{E}_I^{(\infty)}$ can be bounded by the the same expression given in ~\eqref{eq:bound_BM}.
Because $\mathcal{Q}\tilde{\rho}(0) = \tilde{\rho}(0) - \mathcal{P}\tilde{\rho}(0)$, the error~\eqref{eq:inhomogeneous_error_2} can be bounded by the triangle inequality
\begin{equation}
        \label{eq:inhomogeneous_error_separate}
        \mathcal{E}^{(\infty)}_I \leq \norm{\sum_{n=1}^{\infty} \mathcal{I}_n(t)\tilde{\rho}(0) - \sum_{n=1}^{2} \mathcal{I}_n(t)\tilde{\rho}(0)}_1 + \norm{\sum_{n=1}^{\infty} \mathcal{I}_n(t)\mathcal{P}\tilde{\rho}(0) - \sum_{n=1}^{2} \mathcal{I}_n(t)\mathcal{P}\tilde{\rho}(0)}_1 \ ,
\end{equation}
where the second term automatically satisfies the bound given in Eq.~\eqref{eq:bound_BM}. To bound the first term, we first write out $\tilde{\rho}(0)$ explicitly for the case of a single system qubit and $H_\mathrm{I} = g\sigma^z\otimes \sum_{k} \lambda_{k} (b^\dagger_{k} + b_{k})$:
\begin{subequations}
    \begin{align}
        \label{eq:rho_tilde_0}
        \tilde{\rho}(0) &= U_p^\dagger \rho_\mathrm{S}(0)\otimes \rho_\mathrm{B} U_p \\
        &=\pqty{e^{iD}\dyad{0}+e^{-iD}\dyad{1}} \rho_\mathrm{S}(0) \otimes \rho_\mathrm{B} \pqty{e^{-iD}\dyad{0}+e^{iD}\dyad{1}} \\
        &=\rho_\mathrm{S}^{00} e^{iD}\rho_\mathrm{B}e^{-iD} + \rho_\mathrm{S}^{11}e^{-iD}\rho_\mathrm{B}e^{iD} + \rho_\mathrm{S}^{10}e^{-iD}\rho_\mathrm{B}e^{-iD} + \rho_\mathrm{S}^{01}e^{iD}\rho_\mathrm{B}e^{iD}\ ,
        \label{eq:rho_tilde_0-c}
    \end{align}
\end{subequations}
where  $U_p$ was defined in Eq.~(47) in the main text, and
\begin{equation}
    D = \sum_k \frac{g\lambda_k}{i\omega_k}(b^\dagger_k - b_k) \ , \qquad \rho_\mathrm{S}^{ij} = \mel{i}{\rho_\mathrm{S}(0)}{j}\dyad{i}{j}\ . 
\end{equation}
Using our conjecture, the formal expansion of $\mathcal{I}_n(t)$ is the same as $\mathcal{K}_n(t)$ with the average operation in the rightmost term being replaced by
\begin{equation}
    \label{eq:inhomogeneous_average}
    \expval{\cdot}_{mn} = \Tr_\mathrm{B}\Bqty{\cdot \ \rho^{mn}_\mathrm{B}} \ ,
\end{equation}
where $\rho_\mathrm{B}^{mn} = e^{n iD}\rho_\mathrm{B}e^{m iD}$ and $m,n \in \Bqty{-1,+1}$.
To illustrate this point, let us consider one term in the 4th order expansion $\mathcal{I}_4(t)\tilde{\rho}(0)$:
\begin{equation}
    \int_0^{t}\int_0^{t_1}\int_0^{t_2}\operatorname{Tr}_{\mathrm{B}}\left[\tilde{\mathcal{L}}(t) \tilde{\mathcal{L}}\left(t_{3}\right) \rho_{\mathrm{B}}\right] \operatorname{Tr}_{\mathrm{B}}\left[\tilde{\mathcal{L}}\left(t_{1}\right) \tilde{\mathcal{L}}\left(t_{2}\right) \tilde{\rho}(0)\right] \dd{t_1}\dd{t_2}\dd{t_3}\ .
\end{equation}
After substituting Eq.~\eqref{eq:rho_tilde_0-c} into the above expression and expanding the commutators, one term in the summation is
\begin{equation}
    \label{eq:fpp}
    \mathcal{F}_{++} = \Delta^4\int_0^{t}\int_0^{t_1}\int_0^{t_2} \dd{t_1}\dd{t_2}\dd{t_3} \expval{\xi_+(t)\xi_-(t_3)} \expval{\xi_+(t_1)\xi_-(t_2)}_{++} \sigma^+ \sigma^-\sigma^+ \sigma^- \rho^{01}_\mathrm{S} \ ,
\end{equation}
where we use $+, -$ in the subscript instead of $+1, -1$ to simplify the notation. Thus, the next step is to calculate the average and two point correlation function under the new ``average'' operator~\eqref{eq:inhomogeneous_average}:
\begin{subequations}
    \begin{align}
        \label{eq:inhomogeneous_k1_k2}
        \expval{\xi_\pm(t)}_{mn} &= \Tr_\mathrm{B}\{e^{n i D}U^\dagger_\mathrm{B}(t) \xi_\pm U_\mathrm{B}(t) e^{m i D}\rho_\mathrm{B}\} \\
        \expval{\xi_\alpha(t_1)\xi_\beta(t_2)}_{mn} &=\Tr_\mathrm{B}\{e^{n i D}U^\dagger_\mathrm{B}(t_1) \xi_\alpha U_\mathrm{B}(t_1) U^\dagger_\mathrm{B}(t_2) \xi_\beta U_\mathrm{B}(t_2) e^{m i D}\rho_\mathrm{B}\} \ .
    \end{align}
\end{subequations}
Eq.~\eqref{eq:inhomogeneous_k1_k2} can be explicitly carried out using the following identities:
\begin{subequations}
    \begin{align}
        \xi_{\pm}(t) \equiv U^\dagger_\mathrm{B}(t) \xi_\pm U_\mathrm{B}(t) &= \exp\Bqty{\pm 2i U^\dagger_\mathrm{B}(t)\sum_k\frac{g\lambda_k}{i\omega_k}\pqty{b^\dagger_k - b_k} U_\mathrm{B}(t)} \notag \\
        &= \exp\Bqty{\pm 2i \sum_k \frac{g\lambda_k}{i\omega_k}\pqty{b^\dagger_k e^{i\omega_k t} - b_k e^{-i\omega_k t}}} \\
        \comm{b^\dagger_k e^{i\omega_k t} - b_k e^{-i\omega_k t}}{b^\dagger_{k'} - b_{k'}} &= 2i\sin(\omega_k t)\delta_{kk'} \ ,
    \end{align}
\end{subequations}
and the Baker-Campbell-Hausdorff (BCH) formula
\begin{equation}
    e^X e^Y = e^{X+Y+\frac{1}{2}\comm{X}{Y}+\dots} \ .
\end{equation}
After some algebraic manipulations, we get
\begin{equation}
    \expval{\xi_\pm(t)}_{mn} = e^{\pm4in \sum_k \frac{g^2\lambda^2_k}{\omega_k^2}\sin(\omega_kt)} \expval{\xi_\pm(t)e^{i(n+m)D}} =     \begin{cases}
      f^\pm_n(t)\expval{\xi_\pm(t)} & m \neq n\\
      f^\pm_n(t)\expval{\xi_\pm(t)\xi_l(0)} & m=n, \ l = \sgn(n)
    \end{cases} \ ,
\end{equation}
where $f^\alpha_n(t) = \exp{4i\alpha n \sum_k \frac{g^2\lambda^2_k}{\omega_k^2}\sin(\omega_kt)}$.
For simplicity, we consider only the Ohmic bath case, for which $\expval{\xi_\pm(t)}=0$ and $\expval{\xi_\pm(t)\xi_l(0)}$ decay exponentially for $t>0$. As a result, we can ignore the first order term $\expval{\xi_\pm(t)}_{mn}\approx 0$. The two-point correlation function is
\begin{equation}
     \expval{\xi_\alpha(t_1)\xi_\beta(t_2)}_{mn}=f^\alpha_n(t_1)f^\beta_n(t_2)\expval{\xi_\alpha(t_1)\xi_\beta(t_2)e^{i(n+m)D}} =  \begin{cases}
      f^\alpha_n(t_1)f^\beta_n(t_2)\expval{\xi_\alpha(t_1)\xi_\beta(t_2)} & m \neq n\\
      f^\alpha_n(t_1)f^\beta_n(t_2)\expval{\xi_\alpha(t_1)\xi_\beta(t_2)\xi_l(0)} & m=n, \ l = \sgn(n)
    \end{cases} \ .
\end{equation}
where the second line equals 0 because $\expval{\xi_\pm(t)}=0$ and the noise is Gaussian. The above result can be generalized to the multi-point correlation function
\begin{equation}
    \label{eq:multi_correlation_mn}
    \expval{\xi_{\alpha_1}(t_1)\xi_{\alpha_2}(t_2)\cdots\xi_{\alpha_k}(t_k)}_{mn} =  \begin{cases}
      f^{\alpha_1}_n(t_1)f^{\alpha_2}_n(t_2)\cdots f^{\alpha_k}_n(t_k)\expval{\xi_{\alpha_1}(t_1)\xi_{\alpha_2}(t_2)\cdots\xi_{\alpha_k}(t_k)} & m \neq n\\
      f^{\alpha_1}_n(t_1)f^{\alpha_2}_n(t_2)\cdots f^{\alpha_k}_n(t_k)\expval{\xi_{\alpha_1}(t_1)\xi_{\alpha_2}(t_2)\cdots\xi_{\alpha_k}(t_k)\xi_l(0)} & m=n, \ l = \sgn(n)
    \end{cases} \ .
\end{equation}
For the $m \neq n$ case, the correlation function under $\expval{\cdot}_{mn}$ is the same as the correlation function under $\expval{\cdot}$ up to an additional phase factor.
For the $m = n$ case, the right hand side of Eq.~\eqref{eq:multi_correlation_mn} can be decomposed into two-point correlation functions by Wick’s theorem. After the decomposition, each term would appear inside the integral as
\begin{equation}
    \int_0^{t_1}\dd{t_2}\int_0^{t_2}\dd{t_3}\cdots\int_0^{t_{k-1}}\dd{t_k} \expval{\xi_{\alpha_1}(t_{\alpha_1})\xi_{\alpha_2}(t_{\alpha_2})} \cdots \expval{\xi_{\alpha_s}(t_s)\xi_l(0)} \ .
\end{equation}
Again, because the particular pair $\expval{\xi_{\alpha_s}(t_s)\xi_l(0)}$ decreases exponentially for $t_s > 0$ and thus has approximately zero overlap with other two point correlation functions under the integral, we ignore all these terms.
Finally, we can sum up every term over the indices $m$ and $n$. For example, for terms that have the same form as Eq.~\eqref{eq:fpp}, the summation is
\begin{equation}
    \label{eq:f_all}
    \mathcal{F}_{++} + \mathcal{F}_{+-} + \mathcal{F}_{-+} + \mathcal{F}_{--} = \Delta^4\int_0^{t}\int_0^{t_1}\int_0^{t_2} \dd{t_1}\dd{t_2}\dd{t_3} \expval{\xi_+(t)\xi_-(t_3)} \expval{\xi_+(t_1)\xi_-(t_2)} \sigma^+ \sigma^-\sigma^+ \sigma^- \tilde{o} \ ,
\end{equation}
where $\tilde{o} = \rho_\mathrm{S}^{00}f^+_{-}(t_1)f^-_{-}(t_2)  + \rho_\mathrm{S}^{11}f^+_{+}(t_1)f^-_{+}(t_2)$. Eq.~\eqref{eq:f_all} can be bounded by
\begin{equation}
    \norm{\mathcal{F}_{++} + \mathcal{F}_{+-} + \mathcal{F}_{-+} + \mathcal{F}_{--}}_1 \leq \Delta^4\int_0^{t}\int_0^{t_1}\int_0^{t_2} \dd{t_1}\dd{t_2}\dd{t_3}\abs{\expval{\xi_+(t)\xi_-(t_3)}}\abs{\expval{\xi_+(t_1)\xi_-(t_2)}} \ ,
\end{equation}
where we make use of the fact $|{\tilde{o}}| \leq |\rho_\mathrm{S}^{00}| + |\rho_\mathrm{S}^{11}|= 1$.
Formally applying the same technique to every term after expanding $\mathcal{I}_n(t)\tilde{\rho}(0)$, allows us to show that the first part of Eq.~\eqref{eq:inhomogeneous_error_separate} also satisfies the  error bound given in Eq.~\eqref{eq:bound_BM}.\footnote{In this case we do not have the Markov approximation error and the error due to changing the integration limit. But the bound can still be relaxed to the form of Eq.~\eqref{eq:bound_BM}.}

In conclusion, the error due to truncation of the inhomogeneous parts to second order has the same scaling as the error due to truncation of the homogeneous parts [Eq.~\eqref{eq:bound_BM}]. As a result, the total truncation error can be combined using a single big-$O$ notation:
\begin{equation}
   \|\rho_{\text{true}}(t) -\rho_{\text{PTRE}}(t) \|_1 \le O\left(\frac{\tau_B}{\tau_{SB}}   e^{12t/\tau_{SB}}\right)\text{ln}\left(\frac{\tau_{SB}}{\tau_B} \right)\ + \mathcal{E}_I^{(2)},
   \label{err:PTRE_I}
\end{equation}
We  discuss the timescales in the polaron frame in the next section.
Before proceeding, we remark that:
\begin{enumerate}
    \item The same analysis also applies to the multi-qubit PTRE described in Eq.~(53) of the main text by replacing $\rho^{ij}_\mathrm{S}$ with $\rho_k^{ij}$
    \begin{equation}
        \rho^{ij}_k = Z_k \rho_\mathrm{S}(0) Z_k \ ,
    \end{equation}
    where $k$ is the index for different qubits.
    \item The analysis is also applicable for time-dependent $\Delta(t)$. In this case, the corresponding two-point correlation function is defined as $C^{\alpha \beta}(t_1, t_2) = \Delta(t_1)\Delta(t_2)\expval{\xi_\alpha(t_1)\xi_\beta(t_2)}$, which can be upper bounded by
    \begin{equation}
        \abs{C^{\alpha \beta}(t_1, t_2)} \leq \Delta^2_m\abs{\expval{\xi_\alpha(t_1)\xi_\beta(t_2)}} \ ,
    \end{equation}
    where  $\Delta_m = \max_{t\in[0, t_f]} \Delta(t)$. This introduces a constant $\Delta_m$ in the definition of the timescales.
    \item The error bound we analyzed in this section does not include the error due to ignoring the first and second order inhomogeneous terms $\mathcal{E}_I^{(2)}$ (which is usually the case in the standard PTRE treatment). Numerical studies show that they can be ignored if $\rho_\mathrm{S}(0)$ is diagonal~\cite{jang_theory_2009}. More rigorous error bounding is an interesting topic for future study.
\end{enumerate}

\section*{SUPPLEMENTARY NOTE 3: TIME SCALES}
\label{sec:PTRE_timescale}

The time scales given in Eq. (13) in the main text can be computed directly for the polaron frame two-point correlation function~\eqref{eq:polaron_correlation_ohmic}. Because the error bound in Eq. (14) scales with $1/\tau_{SB}$ and $\tau_B/\tau_{SB}$, we explicitly write out those two quantities
\begin{subequations}
    \begin{align}
        \frac{1}{\tau_{SB}} &= \Delta_m^2\int_0^\infty \abs{K(\tau)} \dd{\tau} =\Delta_m^2\int_0^\infty  \exp{-4 g^2\int_0^\infty \frac{J(\omega)(1-\cos(\omega \tau))}{\omega^2}\coth(\beta\omega/2) \dd{\omega}} \dd{\tau}\\
        \frac{\tau_B}{\tau_{SB}} &= \int_0^{t_f} \tau \abs{K(\tau)} \dd{\tau} = \int_0^\infty  \tau\exp{-4 g^2\int_0^\infty \frac{J(\omega)(1-\cos(\omega \tau))}{\omega^2}\coth(\beta\omega/2) \dd{\omega}} \dd{\tau} \ ,
    \end{align}
\end{subequations}
where $\Delta_m = \max_{t\in[0, t_f]} \Delta(t)$. We can see that, when the system-bath coupling strength is sent to infinity, both of these quantities go to zero: $1/\tau_{SB}\to 0$ and $\tau_B/\tau_{SB} \to 0$ as $g\to\infty$. Thus, the PTRE works in the strong coupling regime.

\section*{SUPPLEMENTARY NOTE 4: LINDBLAD FORM OF THE PTRE VIA THE ADIABATIC
APPROXIMATION}
\label{sec:adiabatic_PTRE}

In this section, we derive an adiabatic form of the PTRE from Eq.~(53) of the main text:
\begin{equation}
\label{eq:app_PTRE}
    \pdv{}{t}\tilde{\rho}_{\mathrm{S}}(t) = -i\comm{\tilde{H}_\mathrm{S}(t)+a(t)\sum_i\kappa_i\sigma_x}{\tilde{\rho}_\mathrm{S}(t)} - \sum_{i} \comm{\sigma_i^\alpha}{\Lambda_i^\alpha(t)\tilde{\rho}_\mathrm{S}(t)} + \text{h.c.} \ ,
\end{equation}
where
\begin{subequations}
    \begin{align}
    \label{eq:app_PTRE_lambda}
    \Lambda_i^\alpha(t) &= a(t)\sum_{\beta} \int_0^{t} a(\tau)K_i^{\alpha\beta}(t - \tau)\tilde{U}_\mathrm{S}(t, \tau)\sigma_i^\beta \tilde{U}_\mathrm{S}^\dagger(t, \tau) \dd{\tau} \\
    K_i^{\alpha\beta}(t - \tau) & =\expval{\xi_i^\alpha(t-\tau) \xi_i^\beta(0)}\\
    \kappa_i &= \expval{\xi_i^{\pm}(t)}\ ,
    \end{align}
\end{subequations}
and
\begin{equation}
    \label{eq:app_polaron_Hamiltonian}
     \tilde{U}_\mathrm{S}(t, \tau) = T_+\exp{-i\int_\tau^t \tilde{H}_\mathrm{S}(\tau') \dd{\tau'}} \ .
\end{equation}
Three approximations need to be applied to obtain the final result:
\subsection*{Markov approximation}
The first step is to move the function $a(t)$ in Eq.~\eqref{eq:app_PTRE_lambda} outside the integral
\begin{equation}
    \int_0^t a(\tau) \cdots \dd{\tau} \to a(t)\int_0^t \cdots \dd{\tau} \ ,
\end{equation}
then perform a change of variable $t-\tau \to \tau$. Eq.~\eqref{eq:app_PTRE_lambda} becomes:
\begin{equation}
    \label{eq:markov_lambda}
    \Lambda_i^\alpha(t) = a^2(t)\sum_{\beta} \int_0^{t} K_i^{\alpha\beta}(\tau)\tilde{U}_\mathrm{S}(t, t-\tau)\sigma_i^\beta \tilde{U}_\mathrm{S}^\dagger(t, t-\tau) \dd{\tau} \ .
\end{equation}
Finally, the integration limit is taken to infinity: $\int_0^{t} \to \int_0^\infty$.

\subsection*{Adiabatic approximation}
A detailed description of the adiabatic approximation is given in Ref.~\cite{albash_quantum_2012}. The core idea is to rewrite the unitary in Eq.~\eqref{eq:markov_lambda} as
\begin{equation}
    \tilde{U}_{\mathrm{S}}(t, t-\tau)=\tilde{U}_\mathrm{S}(t,0)\tilde{U}^\dagger_{\mathrm{S}}(t-\tau, 0) \ ,
\end{equation}
and then replace $\tilde{U}_\mathrm{S}$ with an appropriate adiabatic evolution
\begin{subequations}
    \begin{align}
        \tilde{U}_\mathrm{S}(t-\tau,0) &\approx e^{i\tau \tilde{H}_\mathrm{S}(t)} U^{\mathrm{ad}}_\mathrm{S}(t, 0) \\
        \label{eq:approx_unitary_s}
        \tilde{U}_\mathrm{S}(t,0) &\approx  U^{\mathrm{ad}}_\mathrm{S}(t, 0) \ ,
    \end{align}
\end{subequations}
where $U^{\mathrm{ad}}_\mathrm{S}(t, 0)$ is the ideal adiabatic evolution
\begin{equation}
    U_{\mathrm{S}}^{\mathrm{ad}}\left(t, t^{\prime}\right)=\sum_{a}\left|\varepsilon_{a}(t)\right\rangle\left\langle\varepsilon_{a}\left(t^{\prime}\right)\right| \mathrm{e}^{-\mathrm{i} \mu_{a}\left(t, t^{\prime}\right)} \ ,
\end{equation}
with a phase
\begin{equation}
    \mu_{a}\left(t, t^{\prime}\right)=\int_{t^{\prime}}^{t} \mathrm{d} \tau\left[\varepsilon_{a}(\tau)-\phi_{a}(\tau)\right] \ .
\end{equation}
In the above expressions, $\ket{\varepsilon_{a}(t)}$ is the instantaneous eigenstate of $\tilde{H}_\mathrm{S}(t)$ and $\phi_{a}(t)={i}\left\langle\varepsilon_{a}(t) \mid \dot{\varepsilon}_{a}(t)\right\rangle$ is the Berry connection. After this procedure, Eq.~\eqref{eq:markov_lambda} becomes:
\begin{equation}
    \label{eq:adiabatic_pt_lambda}
    \Lambda_{i}^\alpha(t) = a^2(t)\sum_{\beta,a,b} \Gamma_i^{\alpha\beta}(\omega_{ba})L_{i\beta}^{ab}(t)  \ ,
\end{equation}
where $\omega_{ba} = \varepsilon_b - \varepsilon_a$, and
\begin{equation}
    L_{i\beta}^{ab}(t) = \mel{\varepsilon_a(t)}{\sigma_i^\beta}{\varepsilon_b(t)} \dyad{\varepsilon_a(t)}{\varepsilon_b(t)} \ .
\end{equation}
$\Gamma_i^{\alpha\beta}$ is the one-sided Fourier transform of the correlation function
\begin{equation}
    \Gamma_i^{\alpha\beta}(\omega) = \int_0^{\infty}\dd{t} K_i^{\alpha\beta}(t) e^{i\omega t} \ .
\end{equation}
Substituting Eq.~\eqref{eq:adiabatic_pt_lambda} into Eq.~\eqref{eq:app_PTRE}, we have the one-sided adiabatic PTRE.
\subsection*{Rotating wave approximation(RWA)}
To perform the RWA, we first need to move the one-sided adiabatic PTRE into the interaction picture with respect to $\tilde{H}_\mathrm{S}(t)$ via the approximated unitary in Eq.~\eqref{eq:approx_unitary_s}. The non-Hamiltonian part of Eq.~\eqref{eq:app_PTRE} is transformed into
\begin{align}
    &- \tilde{U}^\dagger_\mathrm{S}(t)\sum_{i} \comm{\sigma_i^\alpha}{\Lambda_i^\alpha(t)\tilde{\rho}_\mathrm{S}(t)}\tilde{U}_\mathrm{S}(t) + h.c. \approx \notag\\ &\quad\sum_{a,b,a',b'} e^{-i\bqty{\mu_{ba}(t,0)+\mu_{b'a'}(t,0)}}a^2(t)\sum_i\Gamma_i^{\alpha\beta}(\omega) \bigg\{  \mel{\varepsilon_a(t)}{\sigma_i^\beta}{\varepsilon_b(t)} \Pi_{ab}(0) \accentset{\approx}{\rho}_\mathrm{S}(t)\Pi_{a'b'}(0)\mel{\varepsilon_{a'}(t)}{\sigma_i^\alpha}{\varepsilon_{b'}(t)} \notag \\
    &\quad - \mel{\varepsilon_{a'}(t)}{\sigma_i^\alpha}{\varepsilon_{b'}(t)} \mel{\varepsilon_a(t)}{\sigma_i^\beta}{\varepsilon_b(t)} \Pi_{ab}(0)\Pi_{a'b'}(0) \accentset{\approx}{\rho}_\mathrm{S}(t)\bigg\} + \text{h.c.} \ ,
\end{align}
where $\Pi_{ab}(0)$ is a shorthand notation for $\dyad{\varepsilon_a(0)}{\varepsilon_b(0)}$. The RWA amounts to keeping only terms where either $a'=b$, $b'=a$ or $a=b$, $a'=b'$. After the fast-oscillating terms are ignored, we move back to the polaron frame and make use of Eq.~(39) of the main text to derive the Lindblad-form PTRE:
\begin{equation}
    \dot{\tilde{\rho}}_\mathrm{S}(t) = -i\comm{\tilde{H}_\mathrm{S}(t)+\tilde{H}_{\mathrm{LS}}(t)+a(t)\sum_i\kappa_i\sigma_x}{\tilde{\rho}_\mathrm{S}(t)} + \sum_{i,\alpha,\beta}\sum_\omega\gamma_i^{\alpha\beta}(\omega)\bqty{L_{i}^{\omega,\beta}(t)\tilde{\rho}_\mathrm{S}(t)L^{\omega,\bar{\alpha}\dagger}_{i}(t)-\frac{1}{2}\acomm{L^{\omega,\bar{\alpha}\dagger}_{i}(t)L_{i}^{\omega,\beta}(t)}{\tilde{\rho}_\mathrm{S}(t)}} \ ,
\end{equation}
where the new Lindblad operators are given by
\begin{equation}
    L_{i}^{\omega,\alpha}(t) = a(t)\sum_{\varepsilon_b - \varepsilon_a = \omega}\mel{\varepsilon_a(t)}{\sigma^\alpha_i}{\varepsilon_b(t)}\dyad{\varepsilon_a(t)}{\varepsilon_b(t)} \ .
\end{equation}
The notation $\bar{\alpha}$ again means the opposite symbol of $\alpha$ in $\Bqty{+, -}$. Since $\sigma_i^{\pm}$ are Hermitian conjugates of each other:
\begin{equation}
    L_i^{-\omega, \alpha}(t) = \bqty{L_i^{\omega. \bar{\alpha}}(t)}^\dagger \ .
\end{equation}
Finally, the Lamb shift term is
\begin{equation}
\tilde{H}_{\mathrm{LS}}(t) = \sum_{i,\alpha,\beta}\sum_{\omega} L_i^{\omega,\bar{\alpha}\dagger}L_i^{\omega,\beta}(t)S_i^{\alpha\beta}(\omega) \ .
\end{equation}
Because the techniques used here are the same as those in \cite{albash_quantum_2012, mozgunov_completely_2020}, the error bounds in these references are still valid.
\section*{SUPPLEMENTARY NOTE 5: HYBRID NOISE}
\label{sec:hybrid_noise}

The two point correlation function in the polaron frame [Eq.~\eqref{eq:app_polaron_correlation}] can be extended to a hybrid noise model~\cite{amin_macroscopic_2008}. The core assumption here is that the bath spectral density can be separated into low frequency and high frequency parts:
\begin{equation}
    \gamma(\omega) = \gamma_L(\omega) + \gamma_H(\omega) \ .
\end{equation}
The integral inside the exponent of Eq.~\eqref{eq:app_polaron_correlation} can also be separated into
\begin{equation}
    \label{eq:freq_separation}
    f_L(t) = \int_{-\infty}^{\infty} \dd{\omega} \gamma_L(\omega) \frac{e^{-i\omega t}-1}{\omega^2} \ ,\quad f_H(t) = \int_{-\infty}^{\infty} \dd{\omega} \gamma_H(\omega) \frac{e^{-i\omega t}-1}{\omega^2} \ .
\end{equation}
If $\gamma_L(\omega)$ is concentrated on low frequencies, a 2nd order Taylor expansion of $e^{-i\omega t}$ is justified. As a result, the first term in Eq.~\eqref{eq:freq_separation} can be written as
\begin{equation}
    f_L(t) = -\frac{1}{2}W^2t^2-i\varepsilon_L t
\end{equation}
where
\begin{equation}
    W^2 = \int \gamma_L(\omega) \dd{\omega} \quad \varepsilon_L = \mathcal{P} \int \frac{\gamma_L(\omega)}{\omega} \dd{\omega} \ ,
\end{equation}
and $\mathcal{P}$ stands for the Cauchy principal value. $W$ and $\varepsilon_L$, usually known as the MRT linewidth and reorganization energy, respectively, are experimentally measurable quantities that are connected through the fluctuation-dissipation theorem~\cite{amin_macroscopic_2008}: $W^2 = 2\varepsilon_L T$. 

Finally, the two point correlation function under this hybrid noise model is
\begin{equation}
    K(t) = \expval{\xi_+(t)\xi_-(0)} = \expval{\xi_-(t)\xi_+(0)} = e^{-4i\varepsilon_L t - 2W^2t^2} \exp{4\int \dd{\omega} \gamma_H(\omega) \frac{e^{-i\omega t}-1}{\omega^2}} \ .
\end{equation}
The corresponding spectral density of $K(t)$ is
\begin{equation}
    \gamma_K(\omega) = \int K(t) e^{i\omega t} \dd{t} = \int \frac{\dd{x}}{2\pi} G_L(\omega-x)G_H(x) \dd{x} \ ,
\end{equation}
where $G_L(\omega)$ and $G_H(\omega)$ are the Fourier transforms of $\exp{4f_L(t)}$ and $\exp{4f_H(t)}$, respectively. The low frequency component is Gaussian:
\begin{equation}
    G_L(\omega) = \int \dd{\omega} e^{-4i\varepsilon_L t - 2W^2t^2} e^{i\omega t}= \sqrt{\frac{\pi}{2W^2}}\exp\Bigg[-\frac{(\omega-4\varepsilon_L)^2}{8W^2}\Bigg] \ ,
\end{equation}
while the high frequency part can be approximated as a single Lorentzian~\cite{smirnov_theory_2018}:
\begin{equation}
    G_H(\omega) = \frac{4\gamma_H(\omega)}{\omega^2 + 4\gamma_H(0)^2} \ .
\end{equation}

Finally, it should be mentioned that the fluctuation-dissipation theorem guarantees that the Kubo-Martin-Schwinger (KMS) condition be satisfied for $G_L(\omega)$. So if $\gamma_H(\omega)$ satisfies KMS condition, the spectral density in the polaron frame also satisfies the KMS condition
\begin{equation}
    \gamma_K(\abs{\omega}) = e^{\beta \abs{\omega}}\gamma_K(-\abs{\omega}) \ .
\end{equation}

\section*{SUPPLEMENTARY NOTE 6: PSEUDOCODE FOR TRUNCATED-SUBSPACE AME}
In this section we present the pseduocode for solving AME in the truncated-subspace. First, we define two sets of grid points $\{t_i\}_{i=1}^N$ and $\{\tau_i\}_{i=1}^{N+1}$ such that
\begin{equation}
    \tau_1 < t_i < \tau_2 < t_2 \cdots \tau_i < t_1 < \tau_{i+1} < \cdots < t_N < \tau_{N+1} \ ,
\end{equation}
and $\tau_1=0$, $\tau_{N+1}=t_f$. The values of $\{t_i\}_{i=1}^N$ are chosen subject to the following constraint
\begin{equation}
    \label{eq:grid_condition}
    \sum_{k'=1}^{l}\big\lvert \langle v_k(\tau_i) \vert v_{k'}(t_{i})\rangle \big\rvert^2 > 1-\epsilon \ ,\quad \forall k\in[1,n] \ ,
\end{equation}
where $\ket{v_k(t_i)}$ is the $k$th eigenstate of the system Hamiltonian $H_\mathrm{S}(t_i)$. Additionally, $l$ is the number of energy levels to use in the simulation, $n$ is the size of the energy subspace where the dynamics is confined and $\epsilon$ is the numerical error tolerance. This condition ensures the lowest $n$ eigenstates of $H_\mathrm{S}(\tau_i)$ can almost be expressed by the lowest $l$ eigenstates of $H_\mathrm{S}(t_{i})$. Then the simulation can be done in a piecewise fashion described in Algorithm \ref{alg:truncated-AME}. In practice, $n$ is highly problem-dependent and needs to be estimated empirically based on the adiabatic theorem.
\begin{algorithm}[H]
    \caption{Truncated-subspace AME}\label{alg:truncated-AME}
    \begin{algorithmic}
    \Require $H(t)=\sum_k f_k(t)M_k$
    \State $\rho_0(\tau_1)=\rho(0)$ \Comment{$\rho(0)$ is the initial state in the computaitonal basis}
    \State $i \gets 1$
    \While{$i \leq N$}
    \State $V_i \gets \mathrm{eigendecomposition\ of} \  H(t_i)$ truncated to lowest $l$ levels \Comment{Each column of $V_i$ corresponds to one eigenvector}
    \State $H_r(t) \gets \sum_k f_k(t) V^\dagger_i M_k V_i$
    \State $A_{n,r} \gets V^\dagger_i A_n V_i$ \Comment{$A_n$ is the nth system-bath coupling operator}
    \State $\rho_r \gets V^\dagger_i \rho_{i-1}(\tau_{i}) V_i$
    \State $\rho_i(t) \gets$ solve AME defined by $H_r$ and $A_{n,r}$ from $\tau_i$ to $\tau_{i+1}$ with initial state $\rho_r$
    \State $i \gets i+1$ 
    \EndWhile
    \end{algorithmic}
    \end{algorithm}

\section*{SUPPLEMENT METHOD 1: ADIABATIC FRAME}
We briefly summarize the adiabatic frame transformation, following Ref.~\cite{chen_why_2020}.
For general multi-qubit annealing, the system Hamiltonian given in Eqs.~\eqref{eq:HS} and~\eqref{eq:Hann} can be formally diagonalized and written as 
\begin{equation}\label{eq:multi_HS}
  H_\mathrm{S}(s) = \sum_n E_n\pqty{s}\dyad{n} \ ,
\end{equation}
where $\Bqty{\ket{n\pqty{s}}}$ is the instantaneous energy eigenbasis [eigenvectors of $H_\mathrm{S}(s)$], and $s=t/t_f$ is the dimensionless instantaneous time. We assume that $H_\mathrm{S}(s)$ is real for all $s$. The system density matrix can be written in the instantaneous energy eigenbasis:
\begin{equation}\label{eq:density_matrix_energy_eigen}
  \rho\pqty{s} = \sum_{nm}\rho_{nm}\dyad{n}{m} \ .
\end{equation}
We call the associated matrix $\tilde{\rho} = \bqty{\rho_{nm}}$ the density matrix in the adiabatic frame. It can shown~\cite{chen_why_2020} that 
\begin{equation}
\dot{\rho}_{nm} = 
    -it_f\pqty{E_n-E_m}\rho_{nm} -i\bqty{-i\sum_{n'\neq n}\braket{n}{\dot{n'}}\rho_{n'm} +i \sum_{m'\neq m}\rho_{nm'}\braket{m'}{\dot{m}}}\ ,
\end{equation}
where the dot denotes differentiation with respect to $s$. I.e., $\tilde{\rho}$ obeys the von Neumann equation $\dot{\tilde{\rho}} = -i\comm{\tilde{H}}{\tilde{\rho}}$, with the effective Hamiltonian
\begin{equation}\label{eq:effective_H}
  \tilde{H} =\begin{pmatrix}
  t_fE_0 & -i \braket{0}{\dot{1}} & \dots\\
  i\braket{0}{\dot{1}} & t_fE_1 & \dots\\
  \vdots& & \ddots
  \end{pmatrix} \ .
\end{equation}
\bibliographystyle{naturemag}
\bibliography{refs.bib}
\end{document}